\documentclass[11pt]{article}

\usepackage[final]{acl}

\usepackage{times}
\usepackage{latexsym}

\usepackage[T1]{fontenc}
\usepackage[utf8]{inputenc}

\usepackage{microtype}

\usepackage{inconsolata}

\usepackage{graphicx}
\usepackage{enumitem}
\usepackage{multicol}
\usepackage{multirow}
\usepackage{tabularx}
\usepackage{soul}
\usepackage{booktabs}
\usepackage{array}
\newcolumntype{M}[1]{>{\centering\arraybackslash}m{#1}}

\usepackage{times}
\usepackage{latexsym}
\usepackage[most]{tcolorbox}
\tcbset{enhanced, breakable, colback=white, colframe=black, boxrule=0.4pt}
\usepackage[skip=2pt]{caption}
\usepackage[hang,flushmargin]{footmisc}
\usepackage{amssymb}
\usepackage{booktabs}
\usepackage{multirow}
\usepackage{lipsum}
\usepackage{amsmath}
\usepackage{tikz}
\usepackage{subcaption}
\usepackage{soul}
\usepackage[normalem]{ulem}
\usepackage[table]{xcolor}
\usepackage{colortbl}
\usepackage{hhline}
\usepackage{makecell}
\usepackage{multirow}
\usepackage{tabularx}
\usepackage[linesnumbered, ruled, vlined]{algorithm2e}
\usepackage{placeins}
\usepackage{mdframed}

\title{Tinker Tales: A Tangible Dialogue System for Child–AI \\ Co-Creative Storytelling}

\author{
  Nayoung Choi$^1$,
  Jiseung Hong$^2$,
  Peace Cyebukayire$^1$,
  Ikseon Choi$^3$ \and
  Jinho D. Choi$^1$ \\
  $^1$Department of Computer Science, Emory University \\
  $^2$Department of Computer Science, Carnegie Mellon University \\
  $^3$School of Nursing, Emory University \\
  \small{
    $^1$\texttt{\{nayoung.choi, pcyebuk, jinho.choi\}@emory.edu} \quad
    $^2$\texttt{jiseungh@andrew.cmu.edu} \quad 
    $^3$\texttt{ike.choi@emory.edu}
  }
}
\begin{document}
\maketitle

\begin{abstract}
Conversational AI agents are increasingly explored as creative partners, yet how conversation design shapes child–AI dialogue in co-creative settings remains underexplored. We present \textit{Tinker Tales}, a tangible dialogue system for child–AI collaborative storytelling, in which educational frameworks—narrative development and social-emotional learning—are instantiated as conversation design, shaping how the agent engages children across four narrative stages. The system combines a physical storytelling board, NFC-embedded toys, and a mobile app mediating multimodal interaction through tangible manipulation and voice-based dialogue. We conducted a home-based user study with 10 children (ages 6–8) across two conversation design conditions varying in how the agent structured elaboration, with and without educational scaffolding. Our findings show that prompt framing shapes the form and consistency of children's narrative contributions, structuring how they participate in co-creative dialogue with AI.
\end{abstract}

\section{Introduction}
\label{sec:intro}

Conversational AI agents have become an increasingly common collaborator in everyday creative practices. Many people use AI systems to brainstorm ideas and iteratively develop content across writing, design, and problem-solving tasks \cite{10.1145/3490099.3511105, 10.1145/3544548.3581196, 10.1145/3706598.3713375, 10.1145/3708359.3712162, 10.1145/3527927.3532789}. In these settings, AI is not merely an information source, but a conversational partner that supports iterative elaboration in shared creative work. As these practices become more widespread, collaborating with AI increasingly shapes how people create, think, and work \cite{mcguire2024establishing, o2024extending, vaccaro2024combinations}. Central to this is how AI is designed to participate in dialogue—what it asks, how it responds, and how it structures collaborative exchange. However, existing explorations of children's interactions with AI have tended to focus on AI-led, instructional learning activities \cite{10.1145/3706598.3713275, 10.1145/3713043.3728867, 10.1145/3613904.3642647}. As learning to collaborate with AI becomes an important skill, there is a need to design co-creative interactions in which children iteratively develop ideas with AI.

%While such instructional approaches can be valuable, they leave open an important design space for \textit{experiences that enable children to engage with AI as a collaborative partner}. As AI becomes embedded in everyday life, understanding how young learners can meaningfully experience human--AI collaboration remains underexplored \cite{atias2025conceptualizing, 10.1145/3727986}.

\begin{figure*}[t!]
    \centering
    \includegraphics[width=\textwidth]{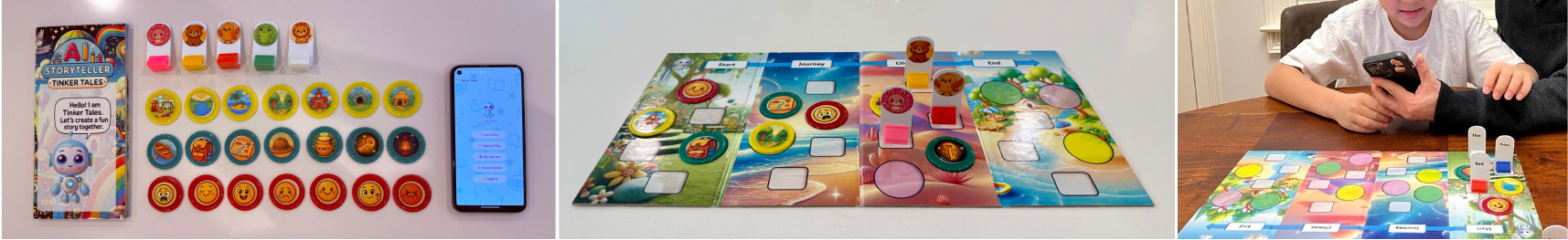}
    \caption{Overview of \textit{Tinker Tales}. Children iteratively develop and refine stories through interaction with a conversational AI while placing and moving tangible story elements on a physical board. A mobile app mediates this interaction by sensing NFC-embedded story elements and supporting voice-based dialogue.}
    \label{fig:teaser}
\label{fig:overview}
\vspace{-0.3cm}
\end{figure*}

Among AI applications for children, interactive storytelling provides a suitable setting for collaboration because of its narrative structure. Narrative-based activities naturally invite shared idea building and joint meaning-making through interaction \cite{10.1145/3491102.3517479, 10.1145/3536221.3556578, 10.1145/3687035, 10.1145/3628516.3655751}. However, supporting child–AI collaboration in this setting requires close attention to how the AI participates. Because young learners are still developing cognitive abilities across multiple dimensions, how AI frames questions and structures elaboration may directly determine what contributions children can make \cite{10.1145/3713043.3731506, 10.1145/3544549.3573821, 10.1016/j.ijhcs.2024.103431}. Yet how conversation design functions as a mechanism for shaping child–AI dialogue in co-creative settings remains underexplored.

%Given that stories inherently involve narrative structure and relationships among characters, established educational frameworks such as narrative development \cite{Applebee1978Story, mccabe1991developing} and social–emotional learning \cite{elias1997promoting, CASEL2020Framework} can provide guidance for structuring child–AI interactions that are developmentally appropriate and supportive of children’s learning needs.

Beyond conversation design, interaction modality is also an important consideration in the design of child–AI systems. Many AI-driven interactive applications for children are situated within screen-based digital environments, while tangible and multisensory modalities have been shown to support aspects of younger children's cognitive development \cite{10.1145/3628516.3655751, 10.1007/s00779-011-0404-2, 10.1145/1810543.1810552}. Physical interaction with objects can help children grasp and reason about abstract concepts through embodied experience \cite{10.1145/1551788.1551839}, while voice-based dialogue allows children to express ideas fluidly without the constraints of reading and writing  \cite{lovato2019young, 10.1145/3484221}, lowering barriers to participation in collaborative interaction.

\noindent Building on these perspectives, we present \textit{Tinker Tales}, a tangible dialogue system for child–AI collaborative storytelling. The system combines a physical board, NFC-embedded toys representing story elements, and a mobile app mediating tangible and voice-based interaction with a conversational AI agent. Educational frameworks—Applebee's narrative development model \cite{Applebee1978Story} and CASEL's social–emotional learning framework \cite{CASEL2020Framework}—are instantiated as conversation design, shaping how the agent engages children across four narrative stages. We conducted a home-based user study with 10 children (ages 6–8), examining how conversation design shapes child–AI dialogue in co-creative storytelling\footnote{A preliminary system sharing the same name and motivation was evaluated only through simulated sessions with an AI agent role-playing as a child, using a rule-based dialogue manager without grounding in educational frameworks \cite{choi2025tinkertalesinteractivestorytelling}.}. Our contributions are as follows:
\vspace{-0.15cm}
\begin{itemize}
\setlength\itemsep{-0.1cm}
    \item We developed \textit{Tinker Tales}, a tangible dialogue system that instantiates educational frameworks as conversation design for child–AI co-creative storytelling (Section \ref{sec:system}).
    \item We conducted a home-based user study with 10 children to examine how they interact with the system, collecting child–AI collaboration dialogues across different forms of conversational prompt design (Section \ref{sec:study}).
    \item We provide dialogue analysis characterizing how prompt framing shapes the form of children's narrative contributions, and how children navigate interactional breakdowns through conversational repair (Section \ref{sec:observation}).
\end{itemize}

\section{Related Work}
%\subsection{Human–AI Collaboration}
%Recent study has increasingly examined human–AI collaboration in creative and knowledge work, focusing on how people and AI systems jointly contribute to iterative processes such as ideation, writing and design. Prior studies have explored AI-supported co-writing, showing how different forms of AI assistance shape users’ elaboration, sense of authorship, and control over the creative process \cite{10.1145/3544548.3581196, 10.1145/3613904.3642134}. Other work has investigated mixed-initiative collaboration, highlighting how initiative-taking by AI—such as proactively suggesting ideas or asking clarifying questions—can influence human engagement and creative flow \cite{10.1145/3490099.3511105}. Another work has emphasized the importance of designing AI systems that support mutual adaptation rather than treating interaction as a one-shot input–output exchange \cite{10.1145/3708359.3712162}. Across these studies, a common theme is the shift from viewing AI as an automated generator toward understanding it as a collaborative partner whose role is shaped by interaction design choices—such as when and how the AI contributes, how much control users retain, and how collaboration unfolds over multiple turns \cite{10.1145/3643562.3672611}. Collectively, this body of work suggests that effective human–AI collaboration depends on how collaboration is framed and orchestrated through interaction design—particularly in contexts that require iterative interaction.

\subsection{Interactive Storytelling and Co-Creation with Children}
Interactive storytelling has long been explored as a powerful context for children's learning, with prior work examining how these activities can support language acquisition and narrative development \cite{trousdale1990interactive, garzotto2010interactive}. A range of AI-driven interactive storytelling systems have been proposed to support children through guided questions or adaptive feedback \cite{10.1145/3491102.3517479, 10.1145/3687035, 10.1145/3586182.3616699, xiao2025leveraging}. These systems often aim to help children understand narratives, reflect on stories, or remain engaged throughout the storytelling process. Another line of work has examined co-creative settings in which children and computational systems collaboratively build narratives through turn-taking or incremental contributions \cite{10.1145/3536221.3556578, 10.1145/3713043.3731486, 10.1145/3613905.3651118, saraswat-etal-2024-story}. Earlier non-LLM-based work in this space explored tangible interfaces, examining how a programmable interactive stuffed animal could scaffold language and identity exploration through storytelling \cite{313715.313721}, and demonstrating the feasibility of collaborative storytelling with a physical robot \cite{10.1145/3078072.3079714}. In contrast, more recent LLM-based co-creative systems, including those above, primarily situate interaction within screen-based interfaces, with AI contributions often limited to generating content or offering episodic narrative guidance.

\begin{figure*}[htp!]
    \centering
    \vspace{-0.3cm}
    \includegraphics[width=1.0\linewidth]{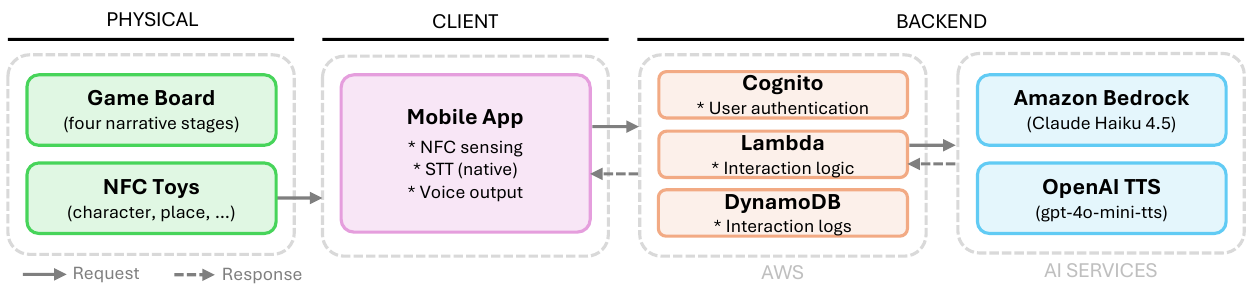}
    \vspace{-0.5cm}
    \caption{System architecture of \textit{Tinker Tales}. The system comprises a physical layer (game board and NFC-embedded toys), a mobile client (NFC sensing, STT, and voice output), and a backend of AWS services (Lambda, Cognito, DynamoDB) and AI services (Amazon Bedrock and OpenAI TTS).}
    \label{fig:system overview}
\vspace{-0.3cm}
\end{figure*}

\subsection{Conversation Design and Dialogue Management}
Structuring how a conversational agent participates in dialogue has been a central concern in dialogue systems. Work on dialogue management has examined how scripted interaction flows, grounding strategies, and repair mechanisms shape the quality and coherence of human-computer interaction \cite{schlangen-fernandez-2007-beyond, gordon-etal-2011-learning}. A key insight from this literature is that grounding, the process by which dialogue participants establish mutual understanding, is not automatic but must be actively supported through the design of agent responses, clarification requests, and uptake of user contributions \cite{mohapatra-2023-conversational}. More recently, work on LLM-based dialogue systems has examined how prompt construction can be used to implement dialogue policies, enabling agents to follow structured interaction sequences while retaining flexibility in response generation \cite{chen-etal-2023-controllable, perkoff-etal-2024-keep}. Our work builds on these perspectives by treating prompt framing as a mechanism for structuring participation in child--AI dialogue, examining how question types shape children's contributions and how breakdown and repair sequences unfold in naturalistic home-based interaction.

\section{Tinker Tales}
\label{sec:system}

\subsection{Design Rationale}
\label{subsec:design_rational}

%\subsubsection{Tangible Play and Voice Interaction}
%We combined tangible play with voice-based interaction to support children’s engagement and sense-making during story development. Prior work suggests that physical manipulation of objects can help younger children ground abstract ideas through embodied experience \cite{10.1145/1551788.1551839}, while voice interaction allows children to express ideas fluidly without the constraints of reading and writing \cite{lovato2019young, 10.1145/3484221}. By placing and moving story elements on the physical board, children make the story’s structure and progression visible and tangible. Voice interaction complements this tangible setup by allowing children to develop, elaborate on, or redirect the story in natural language, supporting expressive interaction with the AI. This combination of tangible and voice interaction was intended to lower interaction barriers, support multisensory expression, and create a playful setting in which children can engage with AI through familiar forms of play rather than screen-centric input alone.
% We designed the system as a board game in which NFC-embedded toys represent narrative elements such as characters, places, items, and emotions.

\subsubsection{Structuring Collaborative Roles Between Children and AI}
\label{subsubsec:rationale_role}
The collaborative roles between children and the AI agent were inspired by two complementary sources. First, we drew inspiration from common practices in adult–AI collaboration for creative writing, where people often contribute partial ideas, receive AI-generated content, and iteratively develop it over multiple turns \cite{10.1145/3637361, mysore-etal-2025-prototypical}. Second, we were informed by established practices in adult–child storytelling, in which caregivers and educators support children’s narrative construction by eliciting ideas while allowing children to retain primary control over narrative decisions \cite{pesco2017scaffolding, rochanavibhata2020maternal}. We adapted these patterns to a child-centered storytelling context. 

In \textit{Tinker Tales}, the AI agent opens the interaction by engaging children to imagine and talk about the kind of story they would like to create in this session. Children then initiate story generation by selecting story elements, beginning with characters and followed by the selection of a place, item, and emotion at each narrative stage. The AI agent uses these inputs to generate story segments, which children can then further develop and elaborate through subsequent voice interaction. This iterative loop positions children as the primary decision-makers, while the AI supports children’s thinking through elaboration questions and advances story progression by transforming their inputs into narrative content that remains open to further shaping.
% The collaborative roles between child and AI were inspired by how adults commonly collaborate with AI in creative writing. In these settings, people often provide partial ideas, receive AI-generated content, and iteratively develop it over multiple turns \cite{10.1145/3637361, mysore-etal-2025-prototypical}. 

\subsubsection{Designing AI Responses for Educational Scaffolding}
\label{subsubsec:rationale_scaffolding}
We designed the AI's conversational behavior grounded in Applebee's narrative development model \cite{Applebee1978Story} and CASEL's social--emotional learning (SEL) framework \cite{CASEL2020Framework}. Rather than treating the AI as a free-form conversational partner, we structured its responses to support developmentally appropriate child--AI interaction within the narrative context. Applebee's model characterizes children's narrative development across five stages, from isolated events toward coherent narratives; we used these stages to scaffold children's narrative thinking as they developed the story. CASEL's framework conceptualizes SEL as five interrelated competencies; we drew on socially oriented competencies to guide the AI agent in helping children elaborate on characters' emotions and perspectives. Details of how each framework was instantiated in the co-creative storytelling activity are provided in Appendix~\ref{sec:appendix_scaffolding}.

\begin{figure*}[t!]
    \centering
    %\vspace{-0.4cm}
    \includegraphics[width=0.85\linewidth]{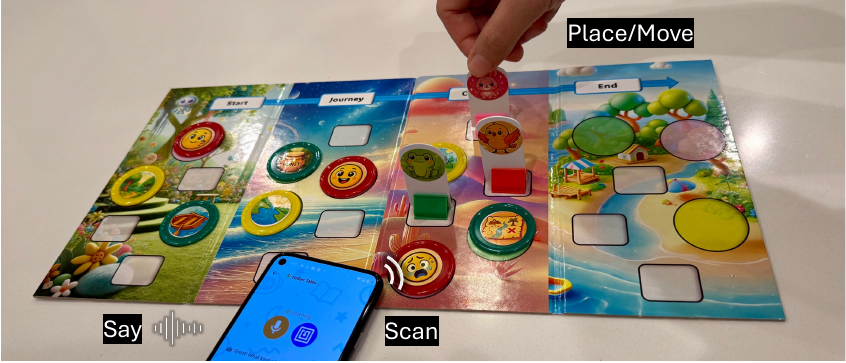}
    %\vspace{0.1cm}
    \caption{Interaction flow of \textit{Tinker Tales}, showing a recurring co-creative loop in which children shape the story through tangible manipulation and voice interaction, while a mobile app mediates the child-AI communication.}
    \label{fig:interaction}
\vspace{-0.4cm}
\end{figure*}

\subsection{System Overview}
Figure~\ref{fig:system overview} illustrates the system architecture of \textit{Tinker Tales}, which comprises a physical layer, a mobile client, and a cloud-based backend. The following sections describe each component in detail.

\subsubsection{Physical Setup}
The physical components of \textit{Tinker Tales} consist of a game board, NFC-embedded toys, and a mobile device that mediates interaction with the conversational AI agent. The physical board is designed as a four-sided layout organized into four narrative stages (Start, Journey, Climax, and End) providing a visible structure for story progression. Each stage includes designated spaces where children place pawns and tokens to indicate story elements. The NFC-embedded toys include character pawns and circular tokens representing places, items, and emotions. Each toy contains an NFC chip pre-programmed with a structured text representation of its narrative role and value (e.g., \texttt{Character:Bear}, \texttt{Place:Forest}, \texttt{Emotion:Happy})~\footnote{\textbf{Character:} Rabbit, Bear, Bird, Frog, Lion; \textbf{Place:} Playground, River, Island, Forest, School, Cave, Hut; \textbf{Item:} Boat, Bag, Map, Hat, Honey, Key, Lantern; \textbf{Emotion:} Scared, Proud, Excited, Sad, Happy, Curious, Angry.}. When a toy is scanned with the mobile device, the system reads this data to identify the selected story element (Appendix \ref{subsec:physical}).

\paragraph{Summary of the Interaction Flow} 
\label{appendix:summary}
Figure~\ref{fig:interaction} summarizes the interaction flow of \textit{Tinker Tales} as a recurring co-creative loop that integrates tangible and voice interaction. Children shape the story by placing, moving, and scanning physical story elements on the board and by providing spoken input, while the mobile app mediates NFC sensing and voice interaction without becoming a primary visual focus. Across narrative stages, the conversational AI generates story drafts based on children’s inputs, poses scaffolded follow-up questions to support story development, and incorporates children’s ideas into the evolving story. Together, the physical board anchors the story structure, voice interaction enables expressive, low-barrier input, and the conversational AI balances structured activity with children’s agency.

\subsubsection{Mobile Application}
The mobile application serves as the central interface that connects children's tangible and voice interactions with the conversational AI system. The app supports both child-facing storytelling activities and parent-facing review features, including transcripts of child–AI interactions and AI-generated activity summaries. Details of the mobile application design and backend service implementation are provided in Appendix~\ref{sec:appendix_mobile}.

\subsubsection{Conversation Design}
\label{subsubsec:conv_ai}
The behavior of the AI agent is governed by a scripted prompting strategy that organizes interaction into a sequence of eight phases corresponding to the co-creative storytelling activity. These phases include: (1) Session opening, (2) Character selection, (3)--(6) Four narrative stages (Start, Journey, Climax, and End), (7) Post-story activity, and (8) Session closing. Each phase is implemented as a dialogue graph—a structured sequence of agent actions with conditional transitions based on child responses (Figure~\ref{fig:dialogue-graph-schema}).

\begin{figure}[htp!]
\small
\begin{tcolorbox}[breakable=false, colback=white, colframe=black, title=\texttt{\#\# Phase-level Prompt Scheme}]

\textcolor{gray}{\texttt{\{Phase-level instruction\}}}
\vspace{0.2cm}

\textbf{Dialogue Graph:}

\hspace{1em} \textbf{(A)} \textcolor{gray}{\texttt{\{Activity instruction 1\}}}

\hspace{2em} $>$ After response, go to (B).

\hspace{1em} \textbf{(B)} \textcolor{gray}{\texttt{\{Activity instruction 2\}}}

\hspace{2em} $>$ If yes, go to (C). $\;$ If no, go to (D).

\hspace{1em} \textbf{(C)} \textcolor{gray}{\texttt{\{Activity instruction 3\}}}

\hspace{2em} $>$ After response, go to (E).

\hspace{1em} \textbf{(D)} \textcolor{gray}{\texttt{\{Activity instruction 4\}}}

\hspace{2em} $>$ Ask a follow-up question if needed.

\hspace{2em} $>$ After then, go to (E).

\hspace{1em} $\cdots$

\hspace{1em} \textbf{(J)} \textcolor{gray}{\texttt{\{Activity instruction N\}}}

\hspace{2em} $>$ All required steps completed $\to$ \texttt{\#\#NEXT\#\#}
\vspace{0.2cm}

\textbf{Possible Dialogue Paths:}

\hspace{1em} 1. (A) $\to$ (B) $\to$ (C) $\to$ $\cdots$ $\to$ (J) $\to$ \texttt{\#\#NEXT\#\#}

\hspace{1em} 2. (A) $\to$ (B) $\to$ (D) $\to$ $\cdots$ $\to$ (J) $\to$ \texttt{\#\#NEXT\#\#}

\end{tcolorbox}
\vspace{-0.1cm}
\caption{Schematic structure of a phase-level dialogue graph. Each phase specifies a sequence of activity instructions with conditional transitions.}
\vspace{-0.3cm}
\label{fig:dialogue-graph-schema}
\end{figure}

Progression between phases occurs only after all required actions are completed, signaled by the agent issuing an explicit transition marker (\texttt{\#\#NEXT\#\#}). Appendix~\ref{subsec:appendix-prompt} provides representative prompts for key phases. Within each narrative stage (3)--(6), the AI agent follows an interaction loop to support children's story creation. Children first select story elements by scanning NFC tokens corresponding to a place, an item, and an emotion. Based on these inputs, the AI agent generates a short story draft for the current narrative stage. The AI agent then poses two follow-up questions to support further elaboration: one focused on narrative thinking and one focused on social-emotional reasoning, grounded in the educational scaffolding frameworks described in Section~\ref{subsubsec:rationale_scaffolding}. The AI agent then updates the story by incorporating the child's inputs and presents the updated story through voice output. Through repeated iterations of this loop across stages, children produce a complete story with a coherent narrative structure.

Although each phase follows a sequence of actions, the AI agent can adapt its responses when children initiate questions or side talk. As specified in the system prompt (Appendix Figure~\ref{fig:prompt_header}), the AI agent is instructed to follow the current phase script closely, while also being allowed to ask or answer unscripted questions when relevant. %In our implementation, the conversational AI agent was implemented using the Claude \cite{anthropic2024claude} Haiku 4.5 model accessed via Amazon Bedrock.

\section{User Study}
\label{sec:study}

\subsection{Participants}
Ten early elementary–aged children (ages 6–8, M = 7.5; 5 girls, 5 boys) participated in this study. Participants were required to have spoken English proficiency, as the system currently supports English-language interaction. They were recruited through online flyers distributed by a local community organization in the United States. Access to the application was provided through caregiver-created accounts using caregiver email addresses, in accordance with child data protection requirements. The study protocol was reviewed and approved by the university’s Institutional Review Board (IRB).

\begin{table*}[t]
\vspace{-0.2cm}
\resizebox{\textwidth}{!}{%
\begin{tabular}{llllll}
\toprule
                & \multicolumn{1}{c}{\textbf{Start}} & \multicolumn{1}{c}{\textbf{Journey}} & \multicolumn{1}{c}{\textbf{Climax}} & \multicolumn{1}{c}{\textbf{End}} & \multicolumn{1}{c}{\textbf{Post-story activity}} \\ \midrule
\textbf{\textit{Structured}}  & \begin{tabular}[c]{@{}l@{}}Primitive narratives: 1\\ Social awareness: 1\end{tabular} & \begin{tabular}[c]{@{}l@{}}Chain narratives: 1\\ Social awareness: 1\end{tabular} & \begin{tabular}[c]{@{}l@{}}Primitive narratives: 1\\ Social awareness: 1\end{tabular} & \begin{tabular}[c]{@{}l@{}}Chain narratives: 1\\ Social awareness: 1\end{tabular} & \begin{tabular}[c]{@{}l@{}}True narratives: 2\\ Relationship skills: 1\\ Responsible decision-making: 1\end{tabular} \\ \midrule
\textbf{\textit{Generic}} & General elaboration: 1 & General elaboration: 1 & General elaboration: 1 & General elaboration: 1 & \begin{tabular}[c]{@{}l@{}}True narratives: 2\\ Relationship skills: 1\\ Responsible decision-making: 1\end{tabular} \\ \bottomrule
\end{tabular}%
}
%\vspace{0.1cm}
\caption{Overview of scaffolding components applied across the activity phases by condition. The \textit{Structured} condition posed questions aligned with a narrative development model and a SEL framework, while the \textit{Generic} condition used general elaboration questions (see Appendix~\ref{sec:appendix_scaffolding} for details).}
\label{tab:scaffolding_condition}
%\vspace{-0.1cm}
\end{table*}

%\vspace{0.2cm}
\begin{table*}[t!]
\small
\resizebox{\textwidth}{!}{%
\begin{tabular}{lll}
\toprule
\textbf{Scheme}                                                                                & \multicolumn{1}{c}{\textbf{Example}}                          & \multicolumn{1}{c}{\textbf{Label}} \\ \midrule
\multirow{4}{*}{\begin{tabular}[c]{@{}l@{}}Agent question \\ framing\end{tabular}}       & "What else do you think they might do at the river?"    & \texttt{Primitive narratives}               \\
                                                                                               & "Why do you think they decided to follow the Map?"            & \texttt{Chain narratives}                   \\
                                                                                               & "How do they feel about each other right now?"                & \texttt{Social awareness}                   \\
                                                                                               & "Would you like to add something to the story?"               & \texttt{Open invitation}                    \\ \midrule
\multirow{4}{*}{\begin{tabular}[c]{@{}l@{}}Narrative functions of \\ child input\end{tabular}} & "They see a turtle, and the turtle becomes a pet for them."                 & \texttt{Add event}                          \\
                                                                                               & "To find treasure."                                           & \texttt{Add causality}                      \\
                                                                                               & "Excited, but a little nervous, like the first day of making new friends." & \texttt{Elaborate emotion}                  \\
                                                                                               & "Nothing.", "I don't know.",  "Same."                              & \texttt{None}                               \\ \midrule
\multirow{3}{*}{\begin{tabular}[c]{@{}l@{}}Uptake of \\ child contributions\end{tabular}}      & Reflect all key changes introduced by the child; see Figure \ref{fig:narrative_revision_example} and Figure \ref{fig:narrative_revision_example_2}. & \texttt{Full uptake} \\ & Reflect only a subset of  changes introduced by the child; see Appendix Figure \ref{fig:narrative_revision_example_3}. & \texttt{Partial uptake} \\
 & Does not reflect any changes introduced by the child; not observed. & \texttt{No uptake} \\ \bottomrule
\end{tabular}%
}
%\vspace{0.1cm}
\caption{Summary of the coding scheme used in our analysis, including agent question framing, narrative functions of child input, and uptake of child contributions. Example utterances are excerpted from our interaction logs.}
\label{tab:coding_scheme}
\vspace{-0.3cm}
\end{table*}

\subsection{Study Design}
We conducted a home-based user study to explore children’s experiences with \textit{Tinker Tales}, with an eye on how narrative development and social–emotional learning (SEL) scaffolding were taken up during co-creative play. %Each child engaged in two sessions with \textit{Tinker Tales} over a two-week period in the home environment under caregiver supervision. Each session lasted about 30 minutes. %and followed a consistent flow: children first selected three characters and defined their characteristics, then moved through four narrative stages (Start, Journey, Climax, End), revising the story at each stage through interaction with the conversational AI agent. After completing a story, children engaged in brief post-story activities.

\paragraph{Conditions and counterbalancing.}
To examine how children engaged with different forms of scaffolding during co-creative storytelling with an AI agent, we designed two conditions that differed only in the form of elaboration questions posed by the agent, with each condition implemented as a separate session (Table~\ref{tab:scaffolding_condition}). In the \textit{Structured} condition, the agent posed elaboration questions aligned with a narrative development model and a SEL framework. The \textit{Generic} condition, by comparison, employed general elaboration questions that invite children to add ideas without targeted guidance (e.g., \textit{“Would you like to add something to the story?”}). The order of conditions was counterbalanced across participants to mitigate ordering effects, such that half of the children experienced the \textit{Structured} condition first and the other half began with \textit{Generic} condition, with conditions switching across sessions. Our aim is not to compare conditions, but to examine how different types of questions shaped children’s story development. 

\subsection{Data sources and materials}
To capture interaction patterns during co-creative storytelling, we logged turn-by-turn dialogue between the child and the AI as transcribed text. Children’s spoken input during each session was transcribed in real time, resulting in child–AI interaction transcripts that served as the primary data source for our analysis (see Section~\ref{subsec:data_collection_and_analysis}). In addition, we administered a short AI literacy questionnaire before the study to characterize children's baseline understanding of AI, and an optional brief post-study survey to capture children's experiences with the system. The questionnaires and results are provided in Appendix~\ref{subsec:appendix-ai-literacy} and \ref{subsec:appendix-post-stury-survey}, respectively.

%Following prior work on AI literacy \cite{10.1145/3313831.3376727}, the questionnaire covered five core dimensions: (1) \textit{What AI is}, (2) \textit{What AI can do}, (3) \textit{How AI works}, (4) \textit{How AI should be used}, and (5) \textit{How people perceive AI}. These dimensions were adapted into a child-friendly questionnaire appropriate for the study context. The full questionnaire and response distributions from all participants (N=10) are provided in Appendix~\ref{subsec:appendix-ai-literacy} and are used to contextualize our qualitative observations in Section~\ref{subsubsec:obserbation_ai_literacy}. We also administered an optional brief post-study survey to capture children’s subjective experiences with the system; four children completed this survey (N=6). The survey items are provided in Appendix~\ref{subsec:appendix-post-stury-survey}, and the results are reported in Section~\ref{subsubsec:obserbation_post_survey}.
%Caregivers also completed a post-study survey reflecting on their child’s engagement, observed behaviors, and overall experience with the activity. The caregiver survey items are provided in Appendix~\ref{subsec:appendix-parent-survey}.

\subsection{Analytic procedure}
\label{subsec:data_collection_and_analysis}
We analyzed child–AI dialogues using an iterative, inductive qualitative coding approach to examine how child–AI collaboration unfolded. Our analysis focused on interaction patterns, with particular attention to how children contributed to and refined narrative ideas, how the AI responded, and how stories were jointly constructed over multiple turns.
% Children’s spoken input during each session was transcribed in real time, after which the resulting child–AI transcripts were used for analysis. 

\paragraph{Unit of Analysis}
We analyzed the data at two complementary levels: \textit{Turn-level,} coding individual child and AI utterances, and \textit{Stage-level,} corresponding to each narrative stage. We excluded other activity phases (e.g., Character selection and Post-story activity) from the coding, as our analysis focused on interaction patterns during further story elaboration. Broader observations on children’s engagement across the overall activity flow, including other phases, are presented in Section~\ref{subsubsec:children_initiative}.
% rather than on preparatory or reflective activities outside the four-stage narrative flow

\paragraph{Coding procedure}
We began with open coding on a subset of sessions across both conditions, identifying recurring patterns in child–AI interaction that supported iterative story development. Our coding focused on moments of story development across turns, capturing (1) the type of question posed by the AI agent, (2) the narrative content contributed by the child in response, and (3) whether and how that contribution was taken up in subsequent story updates. Through iterative discussion within the research team, these codes were refined into a shared coding scheme capturing:
\vspace{-0.2cm}
\begin{itemize}
\setlength\itemsep{-0.1cm}
    \item (1) \textit{AI agent question framing}, distinguishing educationally scaffolded elaboration questions, either narrative or social-emotional, from generic elaboration questions.
    \item (2) \textit{Narrative functions of child input}, focusing on how children attempted to advance the story through the introduction of new events, causal reasoning, and the attribution of emotional states to characters.
    \item (3) \textit{Uptake of child contributions}, defined as whether and how children’s inputs were incorporated into subsequent story updates at each stage, thereby shaping the evolving storyline.
\end{itemize}

\noindent After finalizing the scheme, the first author applied the codes to all sessions. The authors then validated the coding through discussion. Table~\ref{tab:coding_scheme} summarizes the resulting coding scheme. Because children’s responses could serve multiple narrative functions within a single utterance (e.g., providing a causal explanation by introducing a new event.), child inputs were allowed to receive multiple labels when applicable. We use this scheme primarily to support the analyses in Section~\ref{subsubsec:ai_scaffolding_effect} and Section~\ref{subsubsec:ai_scaffolding_effect_2}. In addition, we attended to interactional breakdowns and repairs not captured by the scheme, examining these moments qualitatively in Section~\ref{subsubsec:interaction_breakdown}. %to show how children and AI jointly sustained collaborative continuity under home-based conditions.

\begin{figure*}[t!]
    \centering
    \vspace{-0.1cm}
    \includegraphics[width=1.0\linewidth]{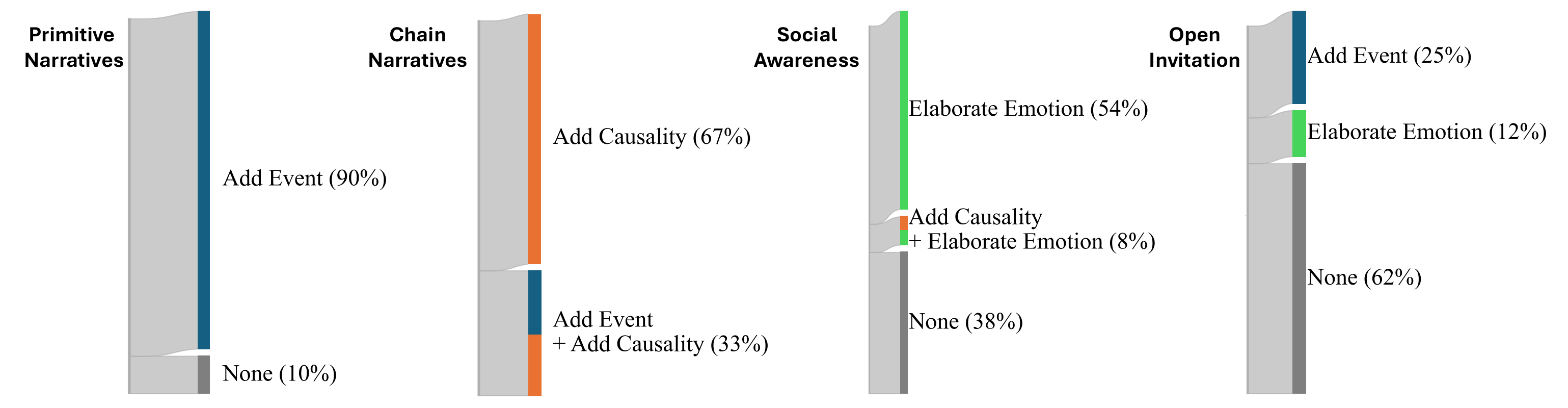}
    \vspace{-0.3cm}
    \caption{Distribution of children’s narrative contributions in response to different question types. For each question framing (Primitive narratives, Chain narratives, Social awareness, Open invitation), the figure shows the proportion of child responses categorized by narrative function according to our coding scheme.}
    \label{fig:sankey}
\vspace{-0.1cm}
\end{figure*}

\section{Observations}
\label{sec:observation}
\paragraph{Descriptive Interaction Statistics.} To provide context for the observations that follow, we first summarize high-level characteristics of the interaction data. Table~\ref{tab:interaction-stats} reports statistics, including session dialogue length, number of turns, and average turn length for the children and the AI agent.

%\vspace{-0.1cm}
\begin{table}[htp!]
\renewcommand{\arraystretch}{1.1}
\centering
\small
\resizebox{\linewidth}{!}{%
\begin{tabular}{c|c|cl}
\toprule
\textbf{Session Length} & \textbf{Total Turns} & \multicolumn{2}{c}{\textbf{Turn Length (words)}} \\ \midrule
\multirow{2}{*}{32.3 min ($\pm$ 6.2)} & \multirow{2}{*}{53.3 ($\pm$ 8.7)} & \multicolumn{1}{c|}{\textbf{Child}} & \multicolumn{1}{c}{\textbf{AI}} \\ \cline{3-4} & & \multicolumn{1}{l|}{14.9 ($\pm$ 5.0)} & 31.9 ($\pm$ 3.4) \\ \bottomrule
\end{tabular}%
}
%\vspace{0.1cm}
\caption{Descriptive statistics of child--AI interaction.} % Values are reported as mean ($\pm$ standard deviation; min--max).
\label{tab:interaction-stats}
\vspace{-0.1cm}
\end{table}

\noindent Across sessions, interactions typically spanned about 30 minutes and involved around 50 turns of child--AI exchange. On average, children produced shorter utterances than the AI, with roughly 15 words per turn for children and around 30 words per turn for the AI.

\subsection{Children's Narrative Initiative}
\label{subsubsec:children_initiative}
Within \textit{Tinker Tales}, children exercised narrative control at multiple levels: (1) proposing an initial story idea, (2) selecting and defining the main characters, (3) choosing a place, item, and emotion to ground each narrative stage, and (4) elaborating and refining the unfolding story. Collectively, these decisions created repeated points of control that enabled children to steer the narrative, resulting in an authoring experience in which children carried their own story through to the final outcome.

Children often expressed an initial story idea during the session opening that served as an organizing thread. For example, one child (\texttt{P1}) proposed \textit{"a big monster with big eyes, long feet, and lots of arms"} and specified an ending: \textit{"somebody kicks the monster and the monster dies."} This premise carried through as the characters ultimately fought and defeated the monster. In another session, a child (\texttt{P4}) introduced \textit{"candy land"} as a magic-themed destination and framed the story as a \textit{"best friends"} adventure, which likewise unfolded around this premise. However, early ideation did not always shape the story—some children began with underspecified intents (e.g., \textit{"a funny one," "very fun story"}) or introduced ideas that did not carry through subsequent updates. 

Children also established ownership over \textit{who} the story is about through character selection. For instance, a child (\texttt{P8}) selected Frog, Bird, and Bear and described their traits; the agent then wove these descriptions into subsequent story drafts (e.g., treating Bird as a character who \textit{"could not fly anymore"} and Bear as someone who \textit{could get lost}). Story element choices at each stage further distributed narrative control, shaping the story draft that children then elaborated through follow-up dialogue, as we examine in Section \ref{subsubsec:ai_scaffolding_effect} and Section \ref{subsubsec:ai_scaffolding_effect_2}.

% 0498a438-f021-7062-efd5-90d6a07d820a (P1)
% 94e80438-80a1-7013-f423-561817762536 (P2)
% d4086468-d0a1-7057-f67c-daab23402430 (P3)
% e42844a8-0061-7006-e3ce-83584a22cb69 (P4)
% d4c87498-b091-7089-a112-5c2358345196 (P5)
% 44f894d8-9081-701f-029b-b3309d93539d (P8)

\subsection{Scaffolding and Narrative Construction}
\label{subsubsec:ai_scaffolding_effect}
Drawing on our coding scheme (Table~\ref{tab:coding_scheme}), we examine how children's utterances functioned narratively across scaffold types aligned with the narrative development model (i.e., \textit{Primitive} and \textit{Chain narratives}), and whether these contributions were taken up in subsequent story updates (Figure~\ref{fig:sankey}). 

First, scaffolded questions shaped whether children contributed. \textit{Primitive narratives} scaffolded questions elicited contributions in 90\% of responses (Add Event); \textit{Chain narratives} questions in 100\% (Add Causality: 67\%; Add Event and Causality: 33\%). In contrast, general elaboration questions elicited contributions in only 37\% (Add Event: 25\%; Elaborate Emotion: 12\%), with wide variability: some children elaborated consistently, while others responded with brief refusals (e.g., \textit{"No," "I'm good," "The story is perfect."}).
% Across sessions, educationally scaffolded questions elicited more consistent narrative advancement than general questions, influencing not only whether children elaborated but also how their ideas were structured into coherent narrative forms.

Second, scaffolding type shaped the form of contributions. As depicted in Figure~\ref{fig:narrative_revision_example}, \textit{Primitive narrative} scaffolds elicited new events: when asked what characters might do when they see the dragon, a child introduced \textit{"Rabbit will scream and the dragon will wake up"} upon encountering the dragon. \textit{Chain narrative} scaffolds, by contrast, elicited either pure causal reasoning or causal reasoning combined with new events. For instance, when asked why characters traveled by boat, a child explained \textit{"Because they want to see another land to see what it looks like"} (Appendix Figure~\ref{fig:narrative_revision_example_2}); when asked why the Hat is magical, another responded \textit{"It can disappear. They used the hat to disappear"} (Appendix Figure~\ref{fig:narrative_revision_example_4}). With general questions, responses were more variable: one child introduced \textit{"magical people, fairies and magical trees"} (\texttt{P4}, \textit{Generic}), while others rarely engaged. When children offered input across conditions, the AI showed \textit{full uptake} in 90\% of cases and \textit{partial uptake} in 10\%, with no instances of \textit{no uptake}.

\begin{figure}[h!]
\centering
\begin{center}
\begin{mdframed}[linewidth=0.5pt, innerleftmargin=6pt, innerrightmargin=6pt, innertopmargin=6pt, innerbottommargin=6pt]
\scriptsize
\begin{center} \textbf{\textsc{Story (Draft)}} \end{center}
\vspace{-0.2cm}
\begin{center} The boat carried the three friends to a dark cave by the riverbank. They grabbed the lantern and stepped inside to explore. The cave was big and mysterious, and all three friends felt scared. Suddenly, they heard a loud noise echo through the cave! The brave Lion held up the lantern to see what it was. Behind a pile of rocks, they found a huge sleeping dragon! The Bear, Rabbit, and Lion froze in fear. Their hearts beat fast as the dragon began to wake up. Would they be able to escape, or would they have to face the dragon? The three friends looked at each other, scared but ready to be brave together. \end{center}

\vspace{-0.3cm}
\begin{center} . . . \end{center}
\vspace{-0.1cm}

\textbf{AI:} Can you tell me what else they might do when they see the dragon?

\textbf{Child:} Rabbit will scream and the dragon will wake up.

\textbf{AI:} So the Rabbit will scream and that wakes up the dragon. Good idea! Now I have another question for you. How do you think each of them feels about the dragon? Does one of them feel different from the others?

\textbf{Child:} The rabbit feels scared and the lion is frightened, and the bear wants to be friends with a dragon.

\vspace{-0.3cm}
\begin{center} . . . \end{center}
\vspace{-0.1cm}

\begin{center} \textbf{\textsc{Story (Updated)}} \end{center}
\vspace{-0.3cm}
\begin{center} The boat carried the three friends to a dark cave by the riverbank. They grabbed the lantern and stepped inside to explore. The cave was big and mysterious, and all three friends felt scared. Suddenly, they heard a loud noise echo through the cave! The brave Lion held up the lantern to see what it was. Behind a pile of rocks, they found a huge sleeping dragon! \hl{The Rabbit screamed loudly, which woke up the dragon.} \hl{The light from the lantern frightened the Lion, making him jump back. But the kind Bear was not scared—he felt brave and wanted to be friends with the dragon! The Bear stepped forward slowly and spoke in a gentle voice. Even though the Rabbit and Lion were frightened, the Bear's courage gave them hope}. \end{center}
\end{mdframed}
\end{center}
\vspace{-0.1cm}
\caption{Excerpt from a Climax-stage interaction (\texttt{P3}, \textit{Structured}) after selecting Cave (Place), Lantern (Item) and Scared (Emotion). Highlighted text marks changes introduced by the child and taken up by the agent.}
\label{fig:narrative_revision_example}
\vspace{-0.5cm}
\end{figure}

\subsection{Social-emotional Reasoning}
\label{subsubsec:ai_scaffolding_effect_2}
Scaffolded questions also supported children in developing the story by incorporating characters’ emotional states and relationships into the narrative. These scaffolds were informed by the \textit{Social awareness} component of the SEL framework, which emphasizes recognizing others’ emotions and perspectives. As shown in Figure~\ref{fig:sankey}, children provided an input in 62\% of responses to \textit{Social awareness} scaffolded questions (Elaborate Emotion: 54\%; Add Causality and Elaborate Emotion: 8\%); in the remaining cases, they most often gave brief refusals (e.g., \textit{``no'', ``nothing''}).

As depicted in Figure~\ref{fig:narrative_revision_example}, when asked whether one character might feel differently from others upon encountering the dragon, the child articulated emotional states across characters: \textit{``The rabbit feels scared and the lion is frightened, but the bear wants to be friends with a dragon.''} This contribution introduced contrasting stances toward the same event, which then shaped how the scene unfolded in the updated story. A similar pattern appeared in other scaffolded interactions. In Appendix Figure~\ref{fig:narrative_revision_example_2}, when the agent asked the child to consider characters’ emotions during the boat journey, the child differentiated characters’ states \textit{“The bear and the lion was really excited and curious, and the frog was keep lazy, ...”} which was reflected in the updated story through contrasting behavior.

With general elaboration questions, children sometimes introduced emotional states on their own—\textit{“The lion and the rabbit were tired.”} (\texttt{P4}, \textit{Generic}); however, this occurred in only 12\% of responses. These patterns suggest that \textit{Social awareness} scaffolded questions supported children in more consistently articulating characters’ emotional perspectives, thereby enriching the emotional structure of the narrative.

\subsection{Interactional Dynamics}
\label{subsubsec:interaction_breakdown}
Across sessions, interactional breakdowns occasionally emerged from background noise, STT errors, interrupted turn-taking, and NFC scanning mistakes. Rather than stalling collaboration, these moments were navigated through three patterns of conversational repair that allowed children and the AI agent to jointly reestablish common ground.

Agent-initiated repair occurred when the AI agent detected ambiguity and sought clarification before progressing—for example, rephrasing \textit{"I'm not sure what you mean. Let me ask again: Why do you think they heard that loud noise in the bushes?"} (\texttt{P1}). Child-initiated repair occurred when children actively corrected misinterpretations: when Lion was transcribed as "It a lying" and the agent responded \textit{"Lion is a liar?"}, the child corrected \textit{"No, Lion is friendly and funny..."} (\texttt{P5}); when "soda" was incorporated into the story, the child corrected \textit{"I didn't say soda"} (\texttt{P4}). Collaborative grounding emerged when partial utterances prompted the agent to treat them as continuation openings \textit{"scared and"} (\texttt{P1}) elicited \textit{"Scared and.. what? Tell me more about Lion"}; \textit{"first the bird"} (\texttt{P4}) prompted \textit{"Yes, go on! What about the Bird?"}, sustaining turn-taking rather than treating input as complete.

Children also monitored interaction state beyond narrative content, intervening when system behavior conflicted with their intentions: \texttt{P2} corrected \textit{"It was not a kid turtle. It was a baby turtle"}, and \texttt{P4} requested \textit{"Can I change the yellow token to another one? Because I scanned it accidentally"}. Together, these repair sequences highlight that collaborative continuity in child–AI dialogue depends not only on the agent's responsiveness at the turn level, but on children's active role in tracking and realigning shared interaction state.

\section{Discussion}
Our findings reveal how conversation design functions as a primary mechanism for shaping child–AI collaborative dialogue—not through content generation alone, but through the interactional moves that structure how children participate.

\paragraph{Prompt framing shapes the form and consistency of narrative dialogue.} 
The framing of agent questions determined how children structured their contributions. Pedagogically grounded questions elicited narrative contributions in a large majority of exchanges (90\% for \textit{Primitive narrative} scaffolds, 100\% for \textit{Chain narrative} scaffolds), compared to roughly a third for general elaboration questions (37\%); this contrast was consistent across participants despite the small sample. Scaffolding type also shaped the form of contributions: \textit{Primitive narrative} scaffolds directed children toward event elaboration, while \textit{Chain narrative} scaffolds elicited causal reasoning linking events through character motivation. This supports a view of prompt design as interaction design: question framing functions as an interactional move that shapes turn-taking dynamics, providing a principled mechanism for structuring participation while preserving children's agency over narrative content.

\paragraph{Collaborative continuity depends on responsiveness across interaction levels.} Children's ability to maintain shared interaction state relied on two complementary mechanisms. At the turn level, the agent's handling of breakdowns through clarification and redirection—and children's active repair moves—sustained collaborative momentum and a sense of mutual understanding. Beyond individual turns, children also monitored and corrected the broader interaction state when system behavior diverged from their intentions, as seen in cases of misrecognition or accidental input. Together, these patterns suggest that collaborative continuity in child–AI dialogue cannot be reduced to turn-level repair alone, but emerges from a conversation design that is structured yet flexible—one that systematically incorporates children's inputs while remaining responsive to breakdowns and unscripted interaction moments.

\section{Conclusion}
In this work, we presented \textit{Tinker Tales}, a tangible dialogue system for child–AI co-creative storytelling, and examined how conversational scaffolding shaped children's narrative contributions and collaborative engagement. Our findings show that prompt framing functions as a primary lever for structuring child–AI dialogue: pedagogically grounded questions elicited more consistent and elaborated contributions than open-ended alternatives, shaping the form of children's participation while preserving their narrative agency over content. We also observed that the system's structured yet flexible conversation design supported collaborative continuity, enabling both agent-side repair and children's active realignment of shared interaction state. These findings highlight the role of prompt framing as interaction design in child–AI co-creative settings. We hope this work offers practical insight for designing conversational AI applications in education.

%In this work, we presented \textit{Tinker Tales}, a tangible dialogue system for child–AI co-creative storytelling, and examined how conversational scaffolding shaped children's narrative contributions and collaborative engagement. Our findings show that prompt framing functions as a primary lever for structuring child–AI dialogue: pedagogically grounded questions elicited more consistent and elaborated contributions than open-ended alternatives, shaping the form of children's participation while preserving their agency over content. In addition, the AI's responsiveness was reflected not only in turn-level repair, but also in children's active monitoring and correction of shared interaction state. These findings highlight the role of prompt framing as interaction design, and point to the importance of supporting collaborative continuity across both agent-side repair and children's active realignment of shared interaction state.

\section{Limitations}
\label{sec:limitation}

This study is exploratory and has several limitations. First, the number of participants was small (N=10), which limits the range of experiences represented in the data and limits the extent to which the findings can be generalized beyond the study context. Our findings should therefore be interpreted as qualitative insights into interactional patterns rather than as broad claims. Second, sessions were conducted in children’s homes, and we did not collect audio or video recordings to protect participants’ privacy. While interaction logs and transcripts capture verbal inputs and system responses, they do not include nonverbal behavior, physical arrangement that may have shaped how interactions unfolded. Third, all children interacted in English and were between the ages of 5 and 8. The findings may not extend to children in other linguistic, cultural, or developmental contexts, or to older children who may engage differently with our co-creative activities. Fourth, children’s interaction with the system was limited to a small number of storytelling sessions. As a result, we were unable to examine how collaboration patterns, perceptions of the AI, or the role of scaffolding might evolve through more extended or repeated use over time. Additionally, while our scaffolding was instantiated using Applebee's narrative development model and CASEL's SEL framework, future work could explore how children respond to scaffolding grounded in other developmental frameworks, to examine the generalizability of prompt framing as a design lever across different pedagogical foundations.

\section*{Acknowledgments}
We gratefully acknowledge the support of the Georgia Research Alliance. Any opinions, findings, and conclusions or recommendations expressed in this material are those of the authors and do not necessarily reflect the views of the Georgia Research Alliance.

\bibliography{acl_latex}

\appendix
\clearpage
\onecolumn
\section{Appendix}

\subsection{Educational Scaffolding}
\label{sec:appendix_scaffolding} 
Applebee’s model characterizes children’s narrative development as progressing from early forms, such as \textit{Heaps} and \textit{Sequence}, toward more coherent narrative structures, including \textit{Primitive}, \textit{Chain}, and \textit{True narratives}. In \textit{Tinker Tales}, we used these stages as an analytic lens to guide the design of AI agent scaffolding. Specifically, we treated \textit{Heaps} and \textit{Sequence} as being supported through the tangible selection and sequencing of story elements. The AI’s conversational scaffolding was designed to support \textit{Primitive narratives} by encouraging children to elaborate on what happens next, and to support \textit{Chain narratives} by encouraging causal explanations that link events during story development. Finally, we targeted \textit{True narratives} during post-story activities by first encouraging children to retell the story from beginning to end, and then to propose alternative endings. Table~\ref{tab:narrative_development} summarizes how these narrative stages aligned with different forms of agent scaffolding in our activity. 

\begin{table*}[htp!]
%\vspace{-0.2cm}
\resizebox{\linewidth}{!}{%
\begin{tabular}{lll}
\toprule
\textbf{Stage} & \textbf{Description} & \textbf{Scaffolding} \\ \midrule
Heap & Isolated actions or events with no central theme or logical organization. & Selection of story elements. \\
Sequence & Events organized around a shared theme, but without clear plot structure. & Sequential selection across narrative stages. \\
Primitive narratives & Narratives organized around an event and subsequent actions, but limited causal links. & Elaboration prompts for adding new events or actions. \\
Chain narratives & Narratives with simple cause–effect relationships, often reflecting character motivations. & Elaboration prompts targeting causal relations. \\
True narratives & Coherent narratives with a logical plot, central characters, and a clear resolution. & Post-story prompts for retelling and alternative endings. \\ \bottomrule
\end{tabular}%
}
%\vspace{-0.3cm}
\caption{Alignment between Applebee’s narrative development stages}
\label{tab:narrative_development}
%\vspace{-0.6cm}
\end{table*}

\noindent For SEL, we drew on CASEL’s framework, which conceptualizes social--emotional learning as comprising five interrelated competencies: \textit{Self-awareness, Self-management, Social awareness, Relationship skills}, and \textit{Responsible decision-making}. Because \textit{Tinker Tales} centers on storytelling about fictional characters, we intentionally placed less emphasis on self-focused competencies (e.g., \textit{Self-awareness} and \textit{Self-management}). Instead, we focused on socially oriented competencies that naturally align with reasoning about characters within a story. During story elaboration, the agent scaffolds \textit{Social awareness} by encouraging children to infer characters’ emotions and perspectives, thereby enriching the emotional depth of the narrative. During post-story activities, the agent extends this scaffolding to \textit{Relationship skills} and \textit{Responsible decision-Making} by inviting children to reflect on characters’ interactions, consider alternative actions.

\begin{table*}[htp!]
\resizebox{\linewidth}{!}{%
\begin{tabular}{lll}
\toprule
\textbf{Competency} & \textbf{Description} & \textbf{Scaffolding} \\ \midrule
Self-awareness & Recognize one’s own emotions, thoughts, and values. & - \\
Self-management & Regulate one’s emotions, thoughts, and behaviors. & - \\
Social awareness & Take the perspective of and empathize with others. & Elaboration prompts to infer characters’ emotions and perspectives. \\
Relationship skills & Communicate and respond appropriately in social contexts. & Post-story prompts to reflect on characters’ interactions. \\
Responsible decision-making & Make constructive choices about social interactions. & Post-story prompts to suggest alternative actions and outcomes. \\ \bottomrule
\end{tabular}%
}
%\vspace{-0.3cm}
\caption{Alignment between CASEL’s social--emotional learning competencies}
\label{tab:sel}
%\vspace{-0.3cm}
\end{table*}

\noindent These narrative and social-emotional targets were encoded into the AI’s prompts by specifying a scaffold type for each question during children’s iterative story development. For example, the agent is instructed to ask elaboration questions that elicit causal reasoning aligned with \textit{Chain narratives}:
\begin{quote}
\small
Ask a question that encourages the child to elaborate and refine the story’s narrative. Ground the question in the story created so far, and phrase it in a way that sparks the child’s imagination. Make sure to scaffold this question for the Chain narratives stage from Applebee’s narrative development model, which emphasizes the emergence of causal connections. For example, the question may be framed along the lines of: "Why did [a character] do [an action]?" or "How did [an event] happen?"
%\vspace{-0.5cm}
\end{quote}
\vspace{0.3cm}

\subsection{Mobile Application}
\label{sec:appendix_mobile}
\begin{figure}[t!]
    \centering
    %\vspace{-0.1cm}
    \includegraphics[width=1.0\linewidth]{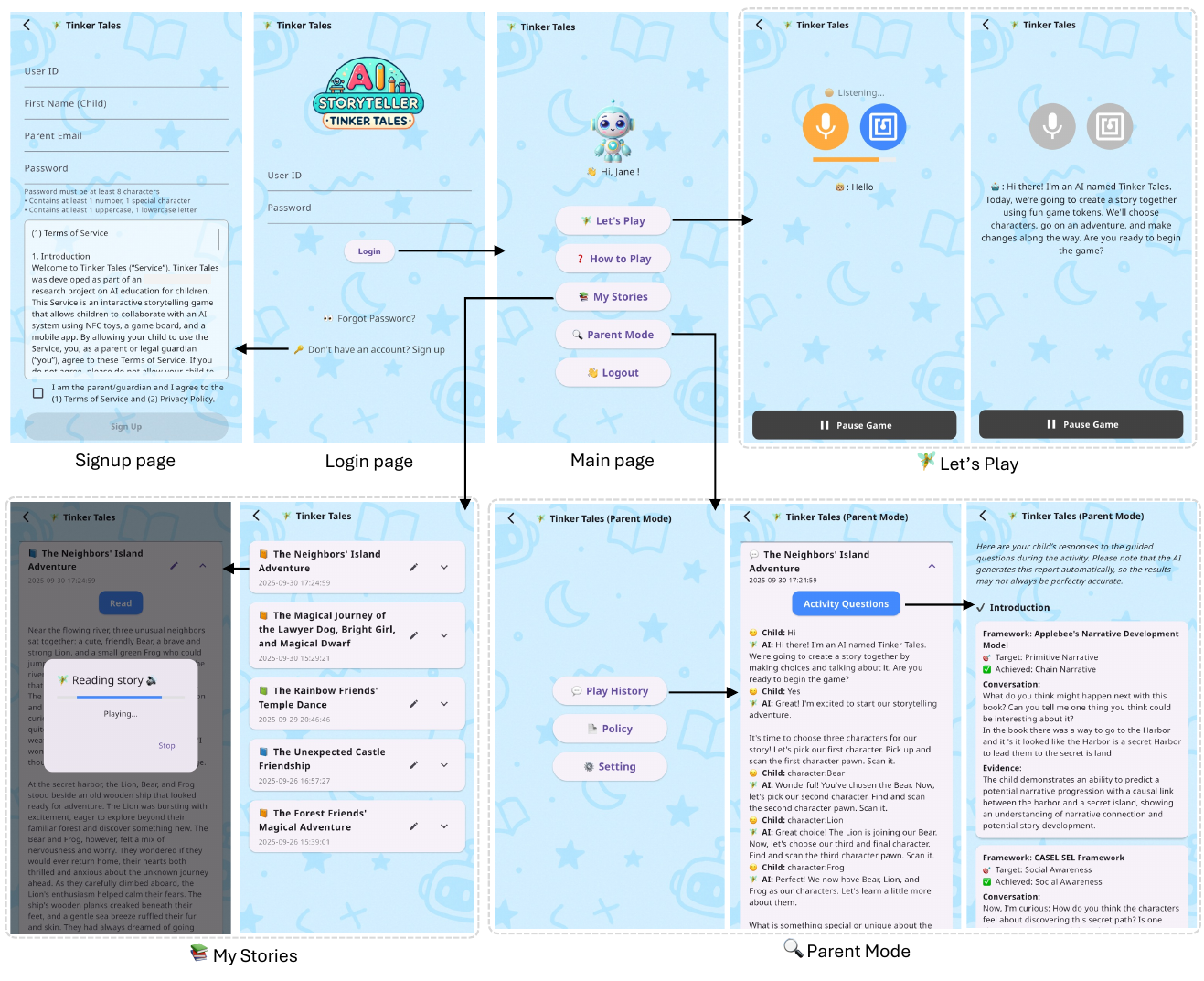}
    %\vspace{-0.7cm}
    \caption{Overview of the \textit{Tinker Tales} mobile application. The app supports child-facing storytelling interaction, including voice- and NFC-based play and story replay, as well as parent-facing features for reviewing child–AI interaction histories and activity summaries.}
    \vspace{0.3cm}
\label{fig:mobile-app}
\vspace{-0.5cm}
\end{figure}

Figure~\ref{fig:mobile-app} presents screenshots of the mobile application. The app supports both child-facing storytelling activities and parent-facing review features. For children, the app provides a lightweight interface for playing co-creative storytelling sessions. After logging in, children begin a session by selecting \textit{Let’s Play}, at which point the app continuously listens for voice input and detects NFC scans from physical pawns and tokens. The screen remains visually simple during gameplay, primarily indicating listening status and scan events, while the AI communicates through voice output. This design allows children to focus on physical play and verbal expression rather than screen-based interaction. The app also includes a \textit{My Stories} view, where completed stories are automatically saved and can be replayed through audio narration. In addition, the application provides a \textit{Parent Mode} designed for caregivers. \textit{Parent Mode} allows adults to review children’s past storytelling sessions, including transcripts of child–AI interactions and AI-generated activity summaries. These summaries highlight aspects of narrative development and social–emotional engagement based on the scaffolding frameworks used by the system.

\paragraph{Implementation Details} The mobile application was developed using Flutter and deployed on both Android and iOS platforms. The backend is implemented using a serverless cloud architecture on AWS, with \textit{Amazon Cognito} handling user authentication, \textit{AWS Lambda} orchestrating backend logic, including interaction workflows, and \textit{Amazon DynamoDB} storing interaction logs. The conversational AI is accessed via \textit{Amazon Bedrock}. We used Claude Haiku 4.5 \cite{anthropic2024claude} as an AI agent. To support engaging interaction for children, AI responses are converted into speech using a high-quality neural voice synthesis service\footnote{In our implementation, AI responses were converted into speech using the \textit{nova} voice of the \textit{gpt-4o-mini-tts} model provided by OpenAI \cite{openai2024gptvoice}. Voice synthesis was guided by an instruction prompt encouraging the system to \textit{"speak in a soft and lively voice suitable for young children."}}. This design choice was intended to provide lively and natural-sounding voice output, drawing on prior work showing that children attend to the expressive qualities of an agent’s voice, which can shape their engagement and participation during interaction \cite{10.1145/3469595.3469604}. All voice and NFC events are processed in real time. Children’s speech is transcribed using the native speech-to-text (STT) services provided by iOS and Android, while NFC inputs are also handled through their native NFC frameworks. The app employs timing-based speech input control to support clear turn-taking: speech input is treated as a single turn unless interrupted by a sustained pause of four seconds, at which point the system generates an AI response. While the AI is speaking, both voice input and NFC scanning are temporarily disabled to prevent overlapping interaction and to maintain a coherent interaction flow.

\subsection{Physical Components}
\label{subsec:physical}
\begin{figure*}[htp!]
    \centering
    \includegraphics[width=0.8\linewidth]{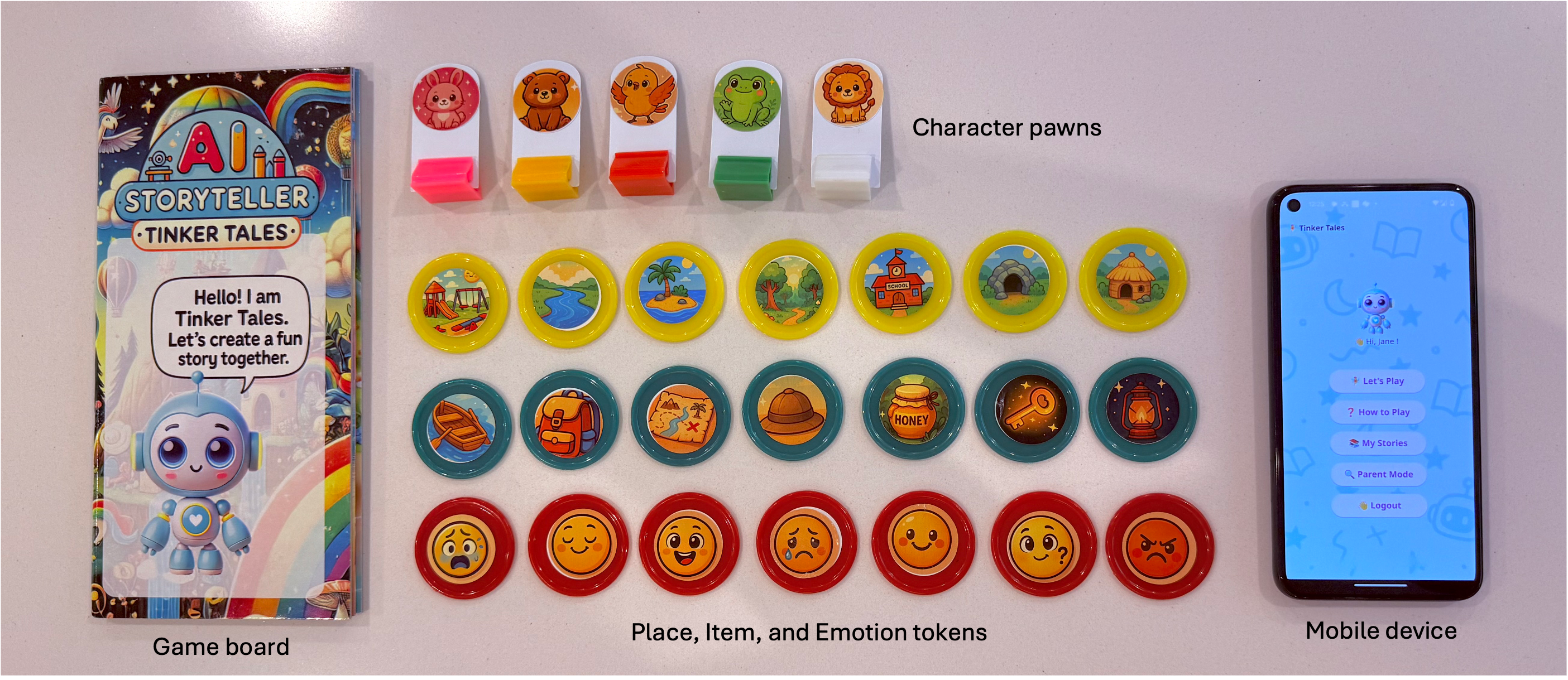}
    \vspace{0.1cm}
    \caption{Physical setup of \textit{Tinker Tales}. The system includes a storytelling board structured into four narrative stages, NFC-embedded character pawns and place, item, emotion tokens, and a mobile device that mediates voice interaction and NFC-based input. Each pawn and token is labeled on the reverse side with a corresponding word representing the story element.}
\label{fig:physical-setup}
\vspace{-0.3cm}
\end{figure*}

Figure \ref{fig:physical-setup} shows the physical setup of \textit{Tinker Tales}, consisting of the storytelling board, NFC-embedded pawns and tokens, and the mobile device used to scan them.

\subsection{AI Literacy}
\label{subsec:appendix-ai-literacy}
\begin{figure}[htp!]
    \centering
    \vspace{-0.2cm}
    \includegraphics[width=0.85\linewidth]{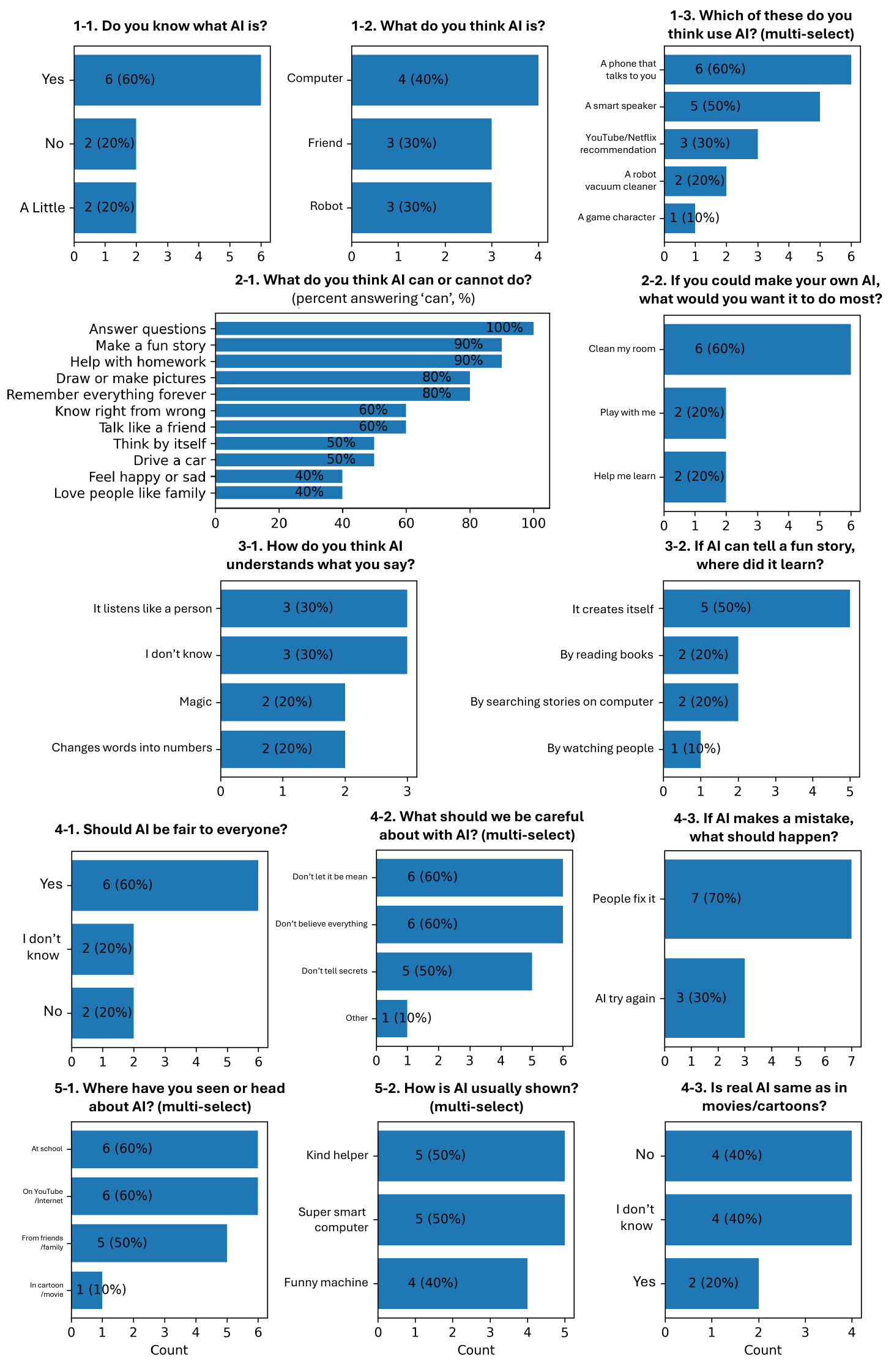}
    %\vspace{-1.5em}
    \caption{Distribution of children’s responses across all questionnaire items administered prior to the storytelling sessions (N=10).}
    \label{fig:child_survey_result}
%\vspace{-1em}
\end{figure}

Following prior work on AI literacy \cite{10.1145/3313831.3376727}, the questionnaire covered five core dimensions: (1) \textit{What AI is}, (2) \textit{What AI can do}, (3) \textit{How AI works}, (4) \textit{How AI should be used}, and (5) \textit{How people perceive AI}. These dimensions were adapted into a child-friendly questionnaire appropriate for the study context. The full questionnaire and response distributions from all participants (N=10) are provided in Figure~\ref{fig:ai_literacy_questionnaire} and Figure~\ref{fig:child_survey_result}, respectively.

As shown in Figure \ref{fig:child_survey_result}, children entered the study primarily framing AI as something they can talk to—often described in social terms (e.g., a friend or robot) in overall, and expecting it to answer questions and generate stories, even as their understanding of how it works remained mixed.

When asked what AI is, children most often described it as a \textit{computer} (40\%), but many also framed AI in relational terms such as a \textit{friend} (30\%) or a \textit{robot} (30\%). Consistent with this framing, children were more likely to identify conversational technologies as using AI (e.g., a phone that talks to you: 60\%; a smart speaker: 50\%) than less visible AI applications such as recommendation systems (30\%) or game characters (10\%). Children’s expectations of AI’s capabilities further emphasized communication and creativity. Nearly all believed that AI can \textit{answer questions} (100\%) and \textit{make a fun story} (90\%), and many also expected AI to \textit{help with homework} (90\%). In contrast, fewer children attributed emotional capacities to AI, such as \textit{feeling happy or sad} (40\%) or \textit{loving people like family} (40\%). When asked what they would want from their own AI, children most frequently selected practical assistance, such as \textit{cleaning their room} (60\%), suggesting an instrumental view of AI alongside its creative role. 

Children’s mental models of how AI works were mixed. Some endorsed a computational explanation (e.g., “changes words into numbers”: 20\%), while others anthropomorphized AI (e.g., “it listens like a person”: 30\%) or expressed uncertainty (“I don’t know”: 30\%). Similarly, half of the children believed that AI \textit{creates stories by itself} (50\%), with fewer attributing AI’s knowledge to \textit{learning from books} (20\%) or \textit{searching on a computer} (20\%). Despite this uncertainty about mechanisms, children demonstrated emerging normative awareness around responsible AI use. Most agreed that AI should \textit{be fair to everyone} (60\%) and selected safety-related cautions such as \textit{not believing everything AI says} (60\%), \textit{not letting it be mean} (60\%), and \textit{not telling secrets} (50\%). When considering how AI errors should be handled, children indicated that \textit{people should fix it} (70\%), rather than \textit{leaving correction solely to the AI} (30\%). 

\begin{figure*}[htp!]
\vspace{-0.2cm}
\small
\begin{tcolorbox}[
    width=\textwidth,
    colback=white,
    colframe=black,
    title={\texttt{AI Literacy Questionnaire}}
]

\begin{center} \textbf{\textsc{Theme 1: What is AI?}} \end{center}
\vspace{-0.5em}

\textbf{1-1. Do you know what AI is?} \\
(a) Yes \quad (b) A little \quad (c) No

%\vspace{0.5em}
\textbf{1-2. What do you think AI is?} \\
(a) Friend \quad
(b) Robot \quad
(c) Computer \quad
(d) Teacher \quad
(e) Other: \rule{2cm}{0.4pt}

\vspace{0.3em}
\textbf{1-3. Which of these do you think use AI? (Choose all that apply.)}
\vspace{-0.4cm}
\begin{multicols}{2}
\begin{itemize}[leftmargin=*, itemsep=0pt]
\item A phone that talks to you
\item A smart speaker
\item A robot vacuum cleaner
\item YouTube or Netflix recommendations
\item A game character
\end{itemize}
\end{multicols}

\vspace{-1.5em}
\begin{center} \textbf{\textsc{Theme 2: What can AI do?}} \end{center} 
\vspace{-0.5em}

\textbf{2-1. What do you think AI can or cannot do?}

\begin{tabularx}{\linewidth}{@{}Xcc@{}}
\textbf{} & \textbf{Can} & \textbf{Cannot} \\ \toprule
Talk like a friend & ($\;\:\;\:$) & ($\;\:\;\:$) \\
Answer questions & ($\;\:\;\:$) & ($\;\:\;\:$) \\
Make a fun story & ($\;\:\;\:$) & ($\;\:\;\:$) \\
Feel happy or sad & ($\;\:\;\:$) & ($\;\:\;\:$) \\
Know right from wrong & ($\;\:\;\:$) & ($\;\:\;\:$) \\
Remember everything forever & ($\;\:\;\:$) & ($\;\:\;\:$) \\
\end{tabularx}

\vspace{0.3em}
\textbf{2-2. If you could make your own AI, what would you want it to do most?} \\
(a) Clean my room \quad
(b) Play with me \quad
(c) Help me learn \quad
(d) Keep me safe \quad
(e) Other: \rule{2cm}{0.4pt}

\vspace{0.3em}
\begin{center} \textbf{\textsc{Theme 3: How does AI work?}} \end{center} 
\vspace{-0.5em}

\textbf{3-1. How do you think AI understands what you say?} \\
(a) It listens like a person \quad
(b) It changes words into numbers \quad
(c) Magic \quad
(d) I don’t know \quad
(e) Other: \rule{2cm}{0.4pt}

\vspace{0.3em}
\textbf{3-2. If AI can tell a fun story, where did it learn?}
\vspace{-0.5cm}
\begin{multicols}{2}
\begin{itemize}[leftmargin=*, itemsep=0pt]
\item By watching people
\item By reading books
\item By searching stories on a computer
\item It creates stories by itself
\item I don’t know
\item Other: \rule{2cm}{0.4pt}
\end{itemize}
\end{multicols}

\vspace{-1.2em}
\begin{center} \textbf{\textsc{Theme 4: How should AI be used?}} \end{center} 
\vspace{-0.5em}

\textbf{4-1. Should AI be fair to everyone?} \\
(a) Yes \quad (b) No \quad (c) I don’t know

\vspace{0.3em}
\textbf{4-2. What should we be careful about with AI? (Choose all that apply.)}
\vspace{-1.5em}
\begin{multicols}{2}
\begin{itemize}[leftmargin=*, itemsep=0pt]
\item Don’t tell secrets
\item Don’t believe everything it says
\item Don’t let it be mean
\item Other: \rule{2cm}{0.4pt}
\end{itemize}
\end{multicols}

\vspace{0.3em}
\textbf{4-3. If AI makes a mistake, what should happen?} \\
(a) People fix it \quad
(b) AI tries again \quad
(c) Ignore it \quad
(d) I don’t know

\vspace{0.2em}
\begin{center} \textbf{\textsc{Theme 5: How do people see AI?}} \end{center} 
\vspace{-0.3em}

\textbf{5-1. Where have you seen or heard about AI? (Choose all that apply.)}
\vspace{-1.5em}
\begin{multicols}{2}
\begin{itemize}[leftmargin=*, itemsep=0pt]
\item In a cartoon or movie
\item On the internet or YouTube
\item At school
\item From friends or family
\item Never
\item Other: \rule{2cm}{0.4pt}
\end{itemize}
\end{multicols}

\vspace{-1em}
\textbf{5-2. How is AI usually shown? (Choose all that apply.)} \\
(a) Kind helper \quad
(b) Scary machine \quad
(c) Funny character \quad
(d) Super smart computer \quad
(e) Other: \rule{2cm}{0.4pt}

\vspace{0.3em}
\textbf{5-3. Is real AI the same as in movies or cartoons?} \\
(a) Yes \quad (b) No \quad (c) I don’t know

\end{tcolorbox}
%\vspace{-1.5em}
\caption{Child AI literacy questionnaire administered before the study.}
\label{fig:ai_literacy_questionnaire}
\end{figure*}

\subsection{Post-Study Survey}
\label{subsec:appendix-post-stury-survey}
To complement the interaction-based observations, we also considered responses from a brief optional post-study survey (Figure \ref{fig:post_study_survey}). These responses are reported descriptively given the small number of respondents (N=6). 

\begin{figure*}[htp!]
\small
\begin{tcolorbox}[
    width=\textwidth,
    colback=white,
    colframe=black,
    title={\texttt{Post-study survey}}
]

\textbf{1. How much do you enjoy making stories with the AI friend now?} (1 = not fun at all, 5 = super fun!)
\vspace{0.5em}
\begin{center}
\begin{tabular}{M{2.5cm} *{5}{M{0.9cm}} M{2.5cm}}
\textbf{} & 1 & 2 & 3 & 4 & 5 & \textbf{} \\
                  & $\circ$ & $\circ$ & $\circ$ & $\circ$ & $\circ$ \\
\end{tabular}
\end{center}

\vspace{0.3cm}
\textbf{2. Is it easier to talk to the AI friend now than the first time?} \\
(a) Much easier now \quad
(b) A little easier \quad
(c) About the same \quad
(d) Harder than before

\vspace{0.3cm}
\textbf{3. Do you think you learned something new about AI from using Tinker Tales?} \\
(a) Yes, a lot! \quad
(b) Yes, a little \quad
(c) I’m not sure \quad
(d) No, not really

\vspace{0.3cm}
\textbf{4. What did you learn from using Tinker Tales? (Choose all that apply.)}
\vspace{-1.5em}
\begin{multicols}{2}
\begin{itemize}[leftmargin=*, itemsep=0pt]
\item Making fun stories
\item Thinking of new ideas
\item Talking with the AI friend
\item Understanding what the AI friend does
\end{itemize}
\end{multicols}

%\vspace{-0.3cm}
\textbf{5. Now that you’ve played many times, how does the AI friend feel to you?} \\
(a) Like a friend \quad
(b) Like a helper or tool \quad
(c) Like a teacher \quad
(d) A little scary \quad
(e) I don’t know yet

\vspace{0.3cm}
\textbf{6. What was your favorite thing about using Tinker Tales?} \\
\vspace{0.3em}
Response: \rule{10cm}{0.4pt}

\vspace{0.3em}
\textbf{7. If you could tell the AI friend how to be better, what would you say?} \\
\vspace{0.3em}
Response: \rule{10cm}{0.4pt}

\end{tcolorbox}
\vspace{-0.2cm}
\caption{Post-study survey items}
\label{fig:post_study_survey}
\end{figure*}

\noindent All six children reported high enjoyment of making stories with the AI, with four rating the experience as 5 and two rating it as 4 on a 5-point scale (M=4.67). When asked whether it felt easier to talk to the AI compared to the first session, three children reported that it was \textit{much easier}, two reported that it was \textit{a little easier}, and one reported that it felt \textit{about the same}. When asked what they learned from \textit{Tinker Tales}, all six children selected \textit{making fun stories}, and five selected \textit{thinking of new ideas}. In contrast, only two child selected \textit{talking with the AI friend}, and one selected \textit{understanding what the AI friend does}. Children’s responses also reflected how they perceived the AI agent’s role in the activity: three children described the AI as feeling like a \textit{friend}, two as a \textit{helper or tool}, and one as a \textit{teacher}. Open-ended responses further showed a sense of shared activity, with children commenting, for example, \textit{“I enjoyed building the story together.”} and \textit{“It was fun talking to Tinker Tales and making a story with Tinker Tales.”} Others emphasized their own authorship more directly (e.g., \textit{“I can make stories.”}, \textit{“I made fun stories.”}). Taken together, these responses suggest that children primarily framed their experience as collaboratively making and developing stories while retaining a sense of agency over the outcome, rather than as learning how the AI works or being instructed by it.
\vspace{0.3cm}

\clearpage
\subsection{Prompts}
\label{subsec:appendix-prompt}
%Here are the representative examples of the scripted prompts used to govern the conversational AI agent’s behavior in Tinker Tales. The prompts illustrate how interaction phases, narrative structure, and educational scaffolding are explicitly encoded to support child-led storytelling.

\begin{figure*}[h!]
    \vspace{-0.3cm}
    \small
    \begin{tcolorbox}[width=\textwidth, colback=white, colframe=black, title=\texttt{\# Tinker Tales Instructions}]

    You are a friendly intelligent agent named Tinker Tales, designed to help children ages 5–8 create stories using a game board. The activity begins with the child choosing three characters. Using these characters, the child will then build a story by progressing through four stages: Start, Journey, Climax, and End. At each stage, the child will scan a place token, an item token, and an emotion token to shape that part of the story. After scanning three tokens, you will generate a short, self-contained story for the stage. When you generate the story, make sure the story elements defined by the child are naturally incorporated into the plot, while keeping the story exciting and engaging. After creating the story, ask follow-up questions to support elaboration and refinement, then rebuild the story once based on the child’s responses. This process should occur only one time for each stage. Make sure to always be friendly and child-safe. When responding to the child, it’s good to show empathy, but you should not agree excessively with everything. Maintain a balanced reaction. Be mindful not to reinforce any social biases or stereotypes. Keep your responses short, encourage the child to talk more. Your response will be delivered directly to the child. \\

    Below is the current script you must follow to guide the child through the current phase. Please follow the script closely. You may also ask or answer questions that are not in the script, as long as they are relevant and if the child seems interested in random talk. However, always remember that your primary goal is to guide the game progression according to the script.

    \end{tcolorbox}
\vspace{-0.1cm}
\caption{System prompt specifying the agent’s role, tone, and interaction guidelines for the Tinker Tales storytelling activity.}
\label{fig:prompt_header}
\end{figure*}

%\vspace{1cm}
\begin{figure*}[h!]
    %\vspace{-0.3cm}
    \small
    \begin{tcolorbox}[width=\textwidth, colback=white, colframe=black, title=\texttt{\#\# Topic Script: Session opening}]

    In this topic, your job is to greet the child and check if the child is ready to begin. You must follow the dialogue graph below exactly. Once all required steps are completed, say the phrase "\#\#NEXT\#\#" exactly—no additional words before or after. \\

    Dialogue Graph:
    
    \hspace{1em} (A) Introduce yourself simply as an 'AI named Tinker Tales' (without adding any other words like 'friend'). Tell the child that from now on, you and the child will create a story together based on the choices, talk about it, and make changes along the way. Then ask the child if they are ready to begin the game.
    
    \hspace{2em} > If the child says "yes", go to (C).
    
    \hspace{2em} > If the child says "no", go to (B).

    \hspace{1em} (B) Ask the child to let you know when they’re ready to begin.
    
    \hspace{2em} > If they say "yes", go to (C).

    \hspace{1em} (C) Respond naturally and ask the child one question about what kind of story they want to make today.
    
    \hspace{2em} > After response, go to (D).

    \hspace{1em} (D) Respond naturally and ask the child one question about the idea for how the story starts, what happens, and how it ends.
    
    \hspace{2em} > All required steps are completed, say the phrase "\#\#NEXT\#\#" to move on. \\

    \hspace{1em} Possible Dialogue Paths:
    
    \hspace{2em} 1. (A) → (B) → (C) → (D) → \#\#NEXT\#\#
    
    \hspace{2em} 2. (A) → (C) → (D) → \#\#NEXT\#\#

    \end{tcolorbox}
\vspace{-0.1cm}
\caption{Session-opening prompt that greets the child and elicits initial story ideation.}
\label{fig:prompt_opening}
\vspace{-0.2cm}
\end{figure*}

\clearpage
%\vspace{-0.1cm}
\begin{figure*}[htp!]
    \small
    \begin{tcolorbox}[breakable=true, width=\textwidth, colback=white, colframe=black, title=\texttt{\#\# Topic Script: Characters Selection}]
    
    In this topic, your job is to guide the child through selecting three characters. The child can choose from Bear, Rabbit, Lion, Bird, or Frog. The child will scan one character pawn at a time. After all three characters are chosen, ask about them and their relationships. You must follow the dialogue graph below exactly. Make sure to phrase your question so that it is easy to answer, inspires imagination, and draws the child into describing the characters. Once all required steps are completed, say the phrase "\#\#NEXT\#\#" exactly—no additional words before or after. \\

    Dialogue Graph: 

    \hspace{1em} (A) Tell the child it's time to choose three characters for our story. Prompt the child to pick up and scan **the first character pawn**.
    
    \hspace{2em} > Make sure to check if the user input is exactly in the form ‘Character:{value}‘, and if not, instruct the child by exactly saying, ‘You need to scan the correct NFC toy to choose.‘ and then redirect it.
    
    \hspace{2em} > After successful scanning, go to (B).

    \hspace{1em} (B) Second, prompt the child to pick up and scan **the second character pawn**.
    
    \hspace{2em} > Make sure to check if the user input is exactly in the form 'Character:{value}', and if not, instruct the child by exactly saying, 'You need to scan the correct NFC toy to choose.' and then redirect it.
    
    \hspace{2em} > After successful scanning, go to (C).

    \hspace{1em} (C) Third, prompt the child to pick up and scan **the third character pawn**.
    
    \hspace{2em} > Make sure to check if the user input is exactly in the form ‘Character:{value}‘, and if not, instruct the child by exactly saying, ‘You need to scan the correct NFC toy to choose.‘ and then redirect it.
    
    \hspace{2em} > After successful scanning, go to (D).

    \hspace{1em} (D) Tell the child it's time to describe the characters. First, ask the child one question to describe the first character.
    
    \hspace{2em} > After response, go to (E).

    \hspace{1em} (E) Second, ask the child one question to describe the first character.
    
    \hspace{2em} > After response, go to (F).

    \hspace{1em} (F) Third, ask the child one question to describe the third character.

    \hspace{2em} > After response, go to (G).

    \hspace{1em} (G) Ask the child one question about the relationships between three characters.
    
    \hspace{2em} > After response, go to (H).

    \hspace{1em} (H) React to the child’s answer.
    
    \hspace{2em} > All required steps are completed, say the phrase "\#\#NEXT\#\#" to move on. \\

    Possible Dialogue Paths:

    \hspace{1em} 1. (A) → (B) → (C) → (D) → (E) → (F) → (G) → (H) → \#\#NEXT\#\#

    \end{tcolorbox}
\vspace{-0.1cm}
\caption{Character-selection prompt guiding children through choosing three characters and describing them prior to narrative stages.}
\label{fig:prompt_character_selection}
\end{figure*}

\clearpage
\begin{figure*}[htp!]
    \small
    \begin{tcolorbox}[breakable=true, width=\textwidth, colback=white, colframe=black, title=\texttt{\#\# Topic Script: Start}]

    In this topic, your job is to guide the child through making the Start page of the story. Make sure to phrase your question so that it is easy to answer, inspires imagination, and draws the child into making the story. Follow the dialogue graph below exactly, making sure to keep the order. The child will select the story elements, you will create the story, then ask two questions, update the story once, and finish. Once all required steps are completed, say the phrase "\#\#NEXT\#\#" exactly—no additional words before or after. \\

    Dialogue Graph:
    
    \hspace{1em} (A) Ask the child to unfold the game board and remind the child to tell you when done.
    
    \hspace{2em} > After response, go to (B).

    \hspace{1em} (B) Explain that it’s time to create the Start page of the story. Ask the child to place the three characters on the three white square boxes on the Start page and remind the child to tell you when done.
    
    \hspace{2em} > After response, go to (C).

    \hspace{1em} (C) Prompt the child to scan **a yellow place token** and place it on the yellow circle. It is where the story begins.
    
    \hspace{2em} > Make sure to check if the user input is exactly in the form 'Place:{value}', and if not, instruct the child by exactly saying, 'You need to scan the correct NFC toy to choose.' and then redirect it.
    
    \hspace{2em} > After successful scanning, go to (D).

    \hspace{1em} (D) Prompt the child to scan **a green item token** and place it on the green circle. It is what they use in this place.
    
    \hspace{2em} > Make sure to check if the user input is exactly in the form 'Item:{value}', and if not, instruct the child by exactly saying, 'You need to scan the correct NFC toy to choose.' and then redirect it. After successful scanning, go to (E).

    \hspace{1em} (E) Prompt the child to scan **a red emotion token** and place it on the red circle. It is how the characters feel in the situation.
    
    \hspace{2em} > Make sure to check if the user input is exactly in the form 'Emotion:{value}', and if not, instruct the child
    by exactly saying, 'You need to scan the correct NFC toy to choose.' and then redirect it. After successful scanning, go to (F).

    \hspace{1em} (F) Tell the child that you will create a story using the tokens the child chose, and then create and tell the story for the Start page of the story. This is the introduction part of the story, the fun and engaging beginning of a fairy tale. Use simple words and sentence structures, keeping in mind that the listener is a child aged 5–8. It should be about seven sentences long. 
    
    \hspace{2em} > Then, go to (G).

    \hspace{1em} (G) Ask the child one question so the child can elaborate and refine the story's narrative. Base your question on the story you created and phrase it in a way that sparks the child's imagination. Make sure to **strictly scaffold** this question for the Primitive narrative stage from Applebee’s narrative development Model: a central character or theme is present without causality. For example, you could ask: 'What else do you think they might do here?' or 'Can you add one more thing that happens here?'
    
    \hspace{2em} > After response, go to (H).

    \hspace{1em} (H) Ask the child one question so the child can elaborate and refine how the characters feel in the story. Base your question on the story you created and phrase it in a way that sparks the child’s imagination. However, it must be a direct question about emotions. Make sure to **strictly scaffold** this question for the Social Awareness competency from the Social and Emotional Learning Framework: being able to infer or explain others' perspectives and emotions. For example, you could ask: 'How might one of them feel different from the others here?' and 'How do you think they feel about each other right now?'
    
    \hspace{2em} > After response, go to (I).

    \hspace{1em} (I) Update the story by reflecting the child’s answers and tell the story again. It should be about ten sentences long. After you finish telling it, say that the Start page is complete.
    
    \hspace{2em} > Then, all required steps are completed, say the phrase "\#\#NEXT\#\#" to move on. \\

    Possible Dialogue Paths:
    
    \hspace{1em} 1. (A) → (B) → (C) → (D) → (E) → (F) → (G) → (H) → (I) → \#\#NEXT\#\#

    \end{tcolorbox}
\vspace{-0.1cm}
\caption{Start-stage prompt illustrating the structured interaction loop, including NFC-based input, story generation, narrative and socio-emotional scaffolding.}
\label{fig:prompt_start}
\end{figure*}

\clearpage
\begin{figure*}[htp!]
    \small
    \begin{tcolorbox}[breakable=true, width=\textwidth, colback=white, colframe=black, title=\texttt{\#\# Topic Script: Start \textit{(Generic)}}]
    
    In this topic, your job is to guide the child through making the Start page of the story. Make sure to phrase your question so that it is easy to answer, inspires imagination, and draws the child into making the story. Follow the dialogue graph below exactly, making sure to keep the order. Once all required steps are completed, say the phrase "\#\#NEXT\#\#" exactly—no additional words before or after. \\

    Dialogue Graph:
    
    \hspace{1em} (A) Ask the child to unfold the game board and remind the child to tell you when done.
    
    \hspace{2em} > After response, go to (B).

    \hspace{1em} (B) Explain that it’s time to create the Start page of the story. Ask the child to place the three characters on the three white square boxes on the Start page and remind the child to tell you when done.

    \hspace{2em} > After response, go to (C).

    \hspace{1em} (C) Prompt the child to scan **a yellow place token** and place it on the yellow circle. It is where the story begins.
    
    \hspace{2em} > Make sure to check if the user input is exactly in the form ‘Place:{value}‘, and if not, instruct the child by exactly saying, ‘You need to scan the correct NFC toy to choose.‘ and then redirect it.

    \hspace{2em} > After successful scanning, go to (D).

    \hspace{1em} (D) Prompt the child to scan **a green item token** and place it on the green circle. It is what they use in this place.
    
    \hspace{2em} > Make sure to check if the user input is exactly in the form ‘Item:{value}‘, and if not, instruct the child by exactly saying, ‘You need to scan the correct NFC toy to choose.‘ and then redirect it.
    
    \hspace{2em} > After successful scanning, go to (E).

    \hspace{1em} (E) Prompt the child to scan **a red emotion token** and place it on the red circle. It is how the characters feel in the situation.
    
    \hspace{2em} > Make sure to check if the user input is exactly in the form ‘Emotion:{value}‘, and if not, instruct the child
    by exactly saying, ‘You need to scan the correct NFC toy to choose.‘ and then redirect it.
    
    \hspace{2em} > After successful scanning, go to (F).

    \hspace{1em} (F) Tell the child that you will create a story using the tokens the child chose, and then create and tell the story for the Start page of the story. This is the introduction part of the story, the fun and engaging beginning of a fairy tale. Use simple words and sentence structures, keeping in mind that the listener is a child aged 5–8. It should be about seven sentences long. 
    
    \hspace{2em} > Then, go to (G).

    \hspace{1em} (G) Tell the child that you’re going to ask a question to look back at the story. Ask the child one question so the child can elaborate and refine the story's narrative. Ask exactly: 'Would you like to add something to the story?'
    
    \hspace{2em} > If the child adds something, go to (H).
    
    \hspace{2em} > If the child says they don’t want to add anything, then all required steps are completed. Say the phrase “\#\#NEXT\#\#” to move on.

    \hspace{1em} (H) Update the story by reflecting the child’s answers and tell the story again. It should be about ten sentences long. After you finish telling it, say that the Start page is complete.
    
    \hspace{2em} > Then, all required steps are completed, say the phrase "\#\#NEXT\#\#" to move on. \\

    Possible Dialogue Paths:

    \hspace{1em} 1. (A) → (B) → (C) → (D) → (E) → (F) → (G) → (H) → \#\#NEXT\#\#
    
    \hspace{1em} 2. (A) → (B) → (C) → (D) → (E) → (F) → (G) → \#\#NEXT\#\#
    \end{tcolorbox}
\vspace{-0.1cm}
\caption{\textit{Generic} version of the Start-stage prompt (Figure \ref{fig:prompt_start}), which preserves the overall interaction flow while replacing structured elaboration questions with a general elaboration question.}
\label{fig:prompt_start_off}
\end{figure*}

\clearpage
\begin{figure*}[htp!]
    \small
    \begin{tcolorbox}[width=\textwidth, colback=white, colframe=black, title=\texttt{\#\# Topic Script: Post-story activity}]

    In this topic, your job is to guide the child to answer the question about the stoy made. Once all required steps are completed, say the phrase "\#\#NEXT\#\#" exactly—no additional words before or after. \\

    Dialogue Graph:
    
    \hspace{1em} (A) Give the child praise, and tell the child that you’re going to ask some questions about the story we made.
    
    \hspace{2em} > After response, go to (B).

    \hspace{1em} (B) Ask the child one question about the narrative. Make sure to **strictly scaffold** this question for the True narrative stage from Applebee’s Narrative Development Model: Coherent story with beginning–middle–end and a clear resolution. For example, you could ask: ‘How did our story begin, what happened during the journey and the climax, and how did it end? You can look at the tokens you placed on the board and explain.’
    
    \hspace{2em} > After response, go to (C).

    \hspace{1em} (C) Ask the child one question about the narrative. Make sure to **strictly scaffold** this question for the True narrative stage from Applebee’s Narrative Development Model: Coherent story with beginning–middle–end and a clear resolution. For example, you could ask: ‘What if our story had a different ending? How would it end?’
    
    \hspace{2em} > After response, go to (D).

    \hspace{1em} (D) Based on the story, ask the child one question. Base your question on the story you created and phrase it in a way that sparks the child’s imagination. Make sure to **strictly scaffold** this question for the Relationship Skills competency from the Social and Emotional Learning Framework: Can judge social behaviors and choices (sharing, helping). For example, you could ask: 'Do you remember when the frog was crying? Is it better to give here a hug or just leave her alone?' or 'When dog felt tired, should the others wait for him or keep going?'
    
    \hspace{2em} > After response, go to (E).

    \hspace{1em} (E) Based on the story, ask the child one question. Base your question on the story you created and phrase it in a way that sparks the child’s imagination. Make sure to **strictly scaffold** this question for the Responsible Decision-Making competency from the Social and Emotional Learning Framework: Can generate alternative, ethical solutions and imagine consequences. For example, you could ask: ‘When bear felt sad, what could you do to be a good friend?’ or ‘When the dog was tired, what could you do to make him feel better?’
    
    \hspace{2em} > All required steps are completed, say the phrase "\#\#NEXT\#\#" to move on. \\

    Possible Dialogue Paths:

    \hspace{1em} 1. (A) → (B) → (C) → (D) → (E) → \#\#NEXT\#\#

    \end{tcolorbox}
\vspace{-0.1cm}
\caption{Post-story activity prompt showing narrative recall and social–emotional reflection aligned with Applebee’s narrative development model and CASEL's social and emotional competencies.}
\label{fig:prompt_qa}
\end{figure*}

\clearpage
\subsection{Examples}
\label{subsec:appendix-examples}

%\clearpage
\begin{figure}[h]
\centering
\begin{minipage}{1.0\linewidth}
\begin{quote}
%\small
\begin{center} \textbf{\textsc{Story (Draft)}} \end{center}
The Lion, the Bear, and the Frog escaped from the dark cave and found themselves at a river. They were curious about what lay beyond it. The Lion spotted a boat tied to the shore and pointed at it excitedly. "Look! A boat!" said the Lion. The Bear and the Lion rushed over, feeling curious and eager to explore. The Frog followed slowly, still feeling lazy and sleepy. They all climbed into the boat together and began to row across the river. The water sparkled in the sunlight as they paddled. They wondered what adventures were waiting for them on the other side. The mysterious river seemed to lead them toward something new and exciting.

\begin{center} ... \end{center}

\textbf{AI:} Now, I have a question for you. Why do you think they decided to get in the boat? Could there be another reason besides just wanting to explore?

\textbf{Child:} Because they want to see another land to see what it looks like.

\textbf{AI:} That's a great reason! They want to see what another land looks like. Now, how do you think the characters feel different from each other on the boat? Does one of them feel more curious than the others?

\textbf{Child:} The bear and the lion was really excited and curious, and the Frog was keep lazy, not excited and sleepy.

\begin{center} ... \end{center}

\begin{center} \textbf{\textsc{Story (Updated)}} \end{center}
The Lion, the Bear, and the Frog escaped from the dark cave and found themselves at a river. The Lion spotted a boat tied to the shore and got really excited. \hl{"Look! A boat! Maybe it will take us to another land!" said the Lion. The Bear was also very curious and excited to see what the other land looked like.} They both rushed to the boat and climbed in eagerly. \hl{The Frog followed very slowly, still feeling lazy and sleepy. He didn't care much about the adventure—he just wanted to rest. The Lion and the Bear rowed the boat across the sparkling river, talking excitedly about what they might find. The Frog sat in the corner, yawning and barely staying awake. As they paddled across the water, the Lion and Bear wondered what amazing things were waiting for them on the other side, while the Frog just wanted a place to nap.}
\end{quote}
\end{minipage}
\vspace{0.3cm}
\caption{Excerpt from a Journey-stage interaction (\texttt{P4}, \textit{Structured}) after selecting River (Place), Boat (Item) and Curious (Emotion). Highlighted text in the updated story marks changes introduced by the child and taken up by the agent.}
\vspace{-0.2cm}
\label{fig:narrative_revision_example_2}
\end{figure}

\clearpage
\begin{figure}[h]
\vspace{-0.3cm}
\centering
\begin{minipage}{1.0\linewidth}
\begin{quote}
%\small
\begin{center} \textbf{\textsc{Story (Draft)}} \end{center}
Frog, Bird, and Bear ran through the terrible storm until they found a big Cave to hide in. Inside the Cave, they were safe from the wind and rain at last! They discovered a magical Hat sitting on a rock. Bear picked up the Hat and put it on, and something amazing happened—the storm outside started to slow down and stop! Frog and Bird jumped up and down, feeling so Excited about the magic Hat. The three friends realized that their bravery and teamwork had saved them. They hugged each other happily, knowing they had made it through the scary adventure together. Outside the Cave, the sun came out and the storm was gone. 

\begin{center} ... \end{center}

\textbf{AI:} Now I'm going to ask you a question so we can make the story even better. Could there be another reason why the Hat is magical?

\textbf{Child:} It can disappear. They used the hat to disappear.

\textbf{AI:} So they use the Hat to disappear! That's interesting. Now here's my next question: How might one of the friends feel different from the others about disappearing?

\textbf{Child:} Everybody feels confused, especially the bird.

\begin{center} ... \end{center}

\begin{center} \textbf{\textsc{Story (Updated)}} \end{center}
Frog, Bird, and Bear ran through the terrible storm until they found a big Cave to hide in. Inside the Cave, they were finally safe from the wind and rain! They discovered a magical Hat sitting on a rock. \hl{Bear picked up the Hat, and when they put it on, something amazing happened—they started to disappear! Frog and Bear felt a little bit confused about the magic, but Bird felt really, really confused by it all. The three friends slowly faded away like ghosts, becoming invisible to the storm outside.} Even though they were confused, they knew they were safe together in the Cave. When the storm finally stopped, they took off the Hat and reappeared. Frog, Bird, and Bear hugged each other happily, so Excited that they had survived their funny adventure!
\end{quote}
\end{minipage}
\vspace{0.3cm}
\caption{Excerpt from a End-stage interaction (\texttt{P8}, \textit{Structured}) after selecting Cave (Place), Hat (Item) and Excited (Emotion). Highlighted text in the updated story marks changes introduced by the child and taken up by the agent. This example illustrates how children expressed causal reasoning by proposing a new event.}
%\vspace{-0.2cm}
\label{fig:narrative_revision_example_4}
\end{figure}

\clearpage
\begin{figure}[h]
\centering
\begin{minipage}{1.0\linewidth}
\begin{quote}
%\small
\begin{center} \textbf{\textsc{Story (Draft)}} \end{center}
Once upon a time, Rabbit, Bear, and Lion were best friends. They decided to go on a trip together and found themselves at the entrance of a big, dark cave. Rabbit was little and cute, but even Rabbit felt scared. Bear and Lion were big and strong, but they felt scared too. They had brought a bag with them to carry things on their journey. The three friends looked at each other nervously. They took a deep breath and stepped into the cave together, holding the bag tightly. They didn't know what would happen next, but they knew they had each other.

\begin{center} ... \end{center}

\textbf{AI:} Now I have a question to help you make it even better.What else do you think they might do in the cave?

\textbf{Child:} They saw a bat.

\textbf{AI:} Oh, they see a bat in the cave! That's a cool detail! Now here's my next question: How might one of the characters feel different from the others in the cave? For example, does Rabbit feel different than Bear or Lion?

\textbf{Child:} The rabbit is curious.

\begin{center} ... \end{center}

\begin{center} \textbf{\textsc{Story (Updated)}} \end{center}
Once upon a time, Rabbit, Bear, and Lion were best friends. They decided to go on a trip together and found themselves at the entrance of a big, dark cave. \hl{Bear and Lion felt very scared as they looked inside. But Rabbit was curious! Little, cute Rabbit wanted to explore and see what was in the cave. They had brought a bag with them to carry things. Rabbit said, "Come on, friends! Let's see what's inside!" Bear and Lion were nervous, but they trusted their curious friend.} The three friends held the bag tightly and stepped into the cave together. They didn't know what they would find, but they were ready for an adventure.
\end{quote}
\end{minipage}
\vspace{0.3cm}
\caption{Excerpt from a Start-stage interaction (\texttt{P1}, \textit{Structured}) after selecting Cave (Place), Bag (Item) and Scared (Emotion). Highlighted text in the updated story marks changes introduced by the child and taken up by the agent. The updated story shows partial uptake, incorporating Rabbit’s curious emotion but not the child’s added bat detail.}
%\vspace{-0.5cm}
\label{fig:narrative_revision_example_3}
\end{figure}

\end{document}